\documentclass{aastex}
\usepackage{spr-astr-addons}
\RequirePackage{color}

\begin{document}
\title{Agegraphic Model based on the Generalized Uncertainty Principle}
\author{S. Davood Sadatian  and  A. Sabouri}
\affil{ Department of Physics, Faculty of Basic Sciences, University
of Neyshabur , P. O. Box 9319774446, Neyshabur, Iran \\
email:sd-sadatian@um.ac.ir}
\begin{abstract}
Many models of dark energy have been proposed to describe the
universe since the beginning of the Big Bang. In this study, we
present a new model of agegraphic dark energy ($NADE$) based on the
three generalized uncertainty principles $KMM$ (Kempf, Mangan,
Mann), Nouicer and $GUP^{*}$ ( higher orders generalized uncertainty
principle).Using the obtained relations from three of types of
$GUP$, in the form of three
scenarios(Emergent,Intermediate,Logamediate), we consider three
different eras of the universe evolution. Also we describe the
evolution and expansion of the universe in each subsection. We will
plot the obtained
relations in these models for better comparatione. 
\end{abstract}

\section{Introduction}
According to Type Ia supernova observations and observed data[1-12],
we know the universe is expanding rapidly today. Extensive studies
show that an energy with unknown nature and negative pressure that
is called dark energy causes this accelerated expansion[13-15]. Dark
energy is one of the most interesting topics in cosmology today, and
various models have been proposed to describe it[16-18]. Only 4\% of
the universe is made of recognizable(known) matter and energy
(baryonic), about 23\% of it is dark matter, and about 73\% is dark
energy, which is now known to cause the universe to expand[19]. In
this regards, many models of dark energy have been proposed to
describe the universe, but so far no one have been able to explain
the nature of dark energy itself. The simplest model for dark energy
is the cosmological constant[20], but this model has a problem with
quantum gravity as the vacuum value of some quantum fields[15]. In
the $\Lambda CDM$ model, dark energy makes about 74\% and dark
matter about 22\% of the universe are so effective on the critical
density of the universe [21-23]. Spergel et al [24] and Seljak et al
[25] combined the data from WMAP and the Supernova Legacy survey to
show that can make a noticeable boundary on state equation of dark
energy in a flat universe($\omega=-0.97_{-0.09}^{+0.07}$). However,
some of the important models that have been proposed for dark energy
are: quintessence[26], phantom[27], quantum[28], dynamical
models[29], tachyon[30], chaplygin gas[31], Ricci[32],
electromagnetic[33], k-essence[34], modified gravity[35] and etc.
There are also two models for dark energy based on the holographic
principle [36], one is the holographic dark energy model [37] and
the other is the agegraphic dark energy model (ADE)[38]. A
holographic model is based on the Bekenstein-Hawking energy boundary
($E_{BH}\geq E_{\Lambda}$) with energy $E_{BH}$ for a universe-size
black hole where $L^{3}\rho_{\Lambda}\leq m_{p}^{2}L$ ($L$ cosmic
length scale, $m_{p}$ Planck mass, $\rho_{\Lambda}$ vacuum energy
density)[39]. Another model based on the karolyhazy relation
[40,41], $\delta t=\lambda t_{p}^{\frac{2}{3}}t^{\frac{1}{3}}$, and
the energy-time uncertainty, $\Delta E\sim t^{-1}$, in Minkowski
space-time which results in $\rho_{q}\sim \frac{\Delta E}{\delta
t^{3}}\sim \frac{m_{p}^{2}}{t^{2}}$ (to estimate the quantum energy
density of Minkowski space-time metric fluctuations)[42]. Here,
$\rho_{q}$ Energy density is called metric perturbations, which is
used with the age of the universe $T=\int_{0}^{t}dt$ as the energy
density of the agrgraphic model. While the vacuum energy density,
$\rho_{\Lambda}$, is used as the energy density of the holographic
model[43,44]. The $ADE$ model considers the relationship between
Heisenberg uncertainty in quantum mechanics and gravitational
effects in general relativity[45]. According to this model, dark
energy is created from fluctuations in space-time and the field of
matter[46]. In the $ADE$ model, instead of the future event horizon,
the age of the universe is used as a measure of length, and this
solves the problem of causality in the holographic model[47]. In
general relativity, space-time can be measured without limitation in
measurement accuracy, but in quantum mechanics the Heisenberg
uncertainty relationship creates measurement accuracy.\\
The original agegraphic model of dark energy ($ADE$) is defined as,
$\rho_{D}=\frac{3n^{2}m_{p}^{2}}{T^{2}}$,($m_{p}=8\pi
G\simeq10^{8}Gev$) which is $G$ Newton's gravity constant, $n^{2}$
represents a constant, and $T$ is the age of the universe ($T=\int
dt=\int\frac{da}{aH}$), which $t$ is cosmic time, $a$ scale factor
and $H$ is a Hubble constant[46]. The basic problem with the $ADE$
model is its contradiction with the matter-dominated era, therefore,
Wei and Cai [38] proposed a new ADE model to solve this problem. For
the age of the universe, they presented $T$ instead of $\eta$, which
was a conformal time scale, and the energy density changed to
$\rho_{D}=\frac{3n^{2}m_{p}^{2}}{\eta^{2}}$. But $\eta$ changed as
$\eta=\int\frac{dt}{a}=\int\frac{da}{a^{2}H}$, i.e. a coefficient
$\frac{1}{a}$ was added to the time scale. Wei and Cai also provided
an important value for the stability of their model, which they
called the squared speed of sound, $v_{s}^{2}$, and defined as
$v_{s}^{2}=\frac{\dot{p}}{\dot{\rho}}$. In order to develop the
stability of the background, the sign of, $v_{s}^{2}$ was very
important, and if $v_{s}^{2}<0$ in general relativity it created an
instability disturbance. In this regard, many models studied such
as: Myung [48] showed that the sign of $v_{s}^{2}$ is negative for
the holographic model, its reason is because of future events
horizon as negative IR cutoff. Kim et al [49] concluded that the
sign of $v_{s}^{2}$ for the agegraphic model is also always negative
and creates instability in the model. Pasqua et al [50] for the dark
energy model based on the uncertainty principle showed that this
model is unstable in the power-law form of the scale factor $a(t)$.
When the effects of quantum gravity become more important, the
Heisenberg uncertainty principle is no longer estimated, and the
generalized uncertainty principle ($GUP$),which is related to string
theory, presents the Planck scale as the minimum length, and hence
describes the evolution of the early universe [51,52]. In the new
agegraphic model($NADE$), two important parameters are the equation
of state, $\omega$, and the square speed of sound, $v_{s}^{2}$,
which $\omega$ determines the nature of the evolution of the
universe and $v_{s}^{2}$ the stability of the evolution of the
universe [49]. Rahul et al [20] investigated a new interactive
agegraphic dark energy model based on $GUP$ and examined the
behavior and evolution of the universe in three scenarios: Emergent,
Intermediate, and Logamediate. In recent years, extensive studies
have been done on the new agegraphic model of dark energy to better
understand the expansion and evolution of the universe like Kurmar
and singh[53] who studied the evolution of the universe at the late
time by the $NADE$ model, and they showed that in this model, by
transition the phase from the matter domination era in the early
universe to an accelerated phase at the late time, can crossed from
the phantom divided line. Also, Hosseinkhani has studied the
thermodynamics of a new interactive agegraphic model [54]. A new
agegraphic model called Tsallis has been proposed by MA Zadeh [55]
and is based on the holographic hypothesis of non-extensive entropy
with an IR cutoff time scale. This work is based on a combination of
articles
[20],[56] and [66].\\
In following, we want to present a new agegraphic dark energy model
($NADE$) in the framework of the three generalized uncertainty
principles mentioned in the reference [56] (
$KMM$,$Nouicer$,$GUP^{*}$) and investigate the evolution and
expansion of the universe in three Emergent, Intermediate,
Logamediate scenarios. In this regard, in section two, we review and
compare the original, new, and interactive agegraphic model, and
then in the third section, we present the new agegraphic model based
on the three $GUPs$ mentioned. Finally, in the fourth section, we
examine the proposed model in the form of three scenarios for the
evolution and expansion of the universe and we summarize our results
in the conclusion section.\\

\section{A review on the agegraphic dark energy models}
In this section, we review and compare the original, new and
interactive agegraphic models of dark energy.
\subsection{Original agegraphic dark energy(ADE)}
As mentioned in the previous section,the basis of the ADE model is
the relation (Karolyhazy) and the energy-time uncertainty in
Minkowski space-time that we have:
\begin{equation} \delta t=\lambda t_{p}^{\frac{2}{3}}t^\frac{1}{3} \end{equation}
\begin{equation} \Delta E\sim t^{-1} \end{equation}
Which is resulted from these two equations:
\begin{equation} \rho_{q}\sim \frac{\Delta E}{\delta t^3}\sim
\frac{m_{p}^{2}}{t^2} \end{equation} In relation $(3)$ instead of
time factor $t$ in this model, the age of the universe is replaced
as follows:[57]
\begin{equation} T=\int_{0}^{a}\frac{da}{aH} \end{equation}
where $a$ is the scale factor and $H\equiv\frac{\dot{a}}{a}$ the
Hubble constant. Energy density in this model is defined as follows:
[57]
\begin{equation} \rho_{q}=\frac{3n^2m_{p}^{2}}{T^2} \end{equation}
where $m_{p}$ is Planck's mass and $T$ is the age of the universe.
In this model, instead of $T$, the age of the universe is set, and
by doing so, the causality problem in the holographic model is
solved and the coefficient $3n^{2}$ is  to parameterize some
uncertainties, such as the effects of space-time curves, various
quantum fields in the universe, and etc [38]. The Friedmann equation
in this model (considering a FRW flat universe containing agegraphic
dark energy and pressureless matter) is defined as follows:
\begin{equation} H^{2}=\frac{1}{3m_{p}^{2}}(\rho_{m}+\rho_{q})
\end{equation}
Using the energy density relationship in this model, the fractional
energy density for pressureless matter and agegraphic dark energy is
equal to:[57]
\begin{equation} \Omega_{q}=\frac{n^{2}}{H^{2}T^{2}} \end{equation}
\begin{equation} \Omega_{m}=1-\Omega_{q} \end{equation}
Using relationships $(5)$,$(6)$,$(7)$ the fractional energy density
for pressureless matter and agegraphic dark energy is equal to:[57]
\begin{equation}
\acute{\Omega_{q}}=\Omega_{q}(1-\Omega_{q})(3-\frac{2\sqrt{\Omega_{q}}}{n})
\end{equation}
Now using the energy conservation equation
$\dot{\rho_{q}}+3H(\rho_{q}+p_{q})=0$ and equations $(5)$ and $(7)$
the equation of parameter of state (EoS) of the agegraphic dark
energy is obtained as:
\begin{equation} \omega_{q}=-1+\frac{2\sqrt{\Omega_{q}}}{3n} \end{equation}
The interesting point about relation $(9)$ is that if
$\Omega_{q}\rightarrow0$ then $w_{q}\rightarrow- 1$ at the early
time while if $\Omega_{q}\rightarrow1$ then $w_{q}\rightarrow -
1+\frac{2}{3n}$ late time. The first is called the matter-dominated
epoch,in which $\acute{\Omega_{q}}\simeq3\Omega{_q}$ and $\omega_{q}
\simeq-1$, as a result $\Omega_{q}\propto a^3$. This means that with
$w_{q}\simeq-1$ the agegraphic dark energy in the epoch of
matter-dominated mimics a cosmic constant while the energy density
of pressureless matter is considered as $p_{m}\propto a^{-3}$[58].
Therefore,there is an implied confusion in this agegraphic dark
energy model[38]. In fact, in the matter-dominated epoch we
have:$$\Omega_{q}\ll1\Rightarrow t^{\frac{2}{3}}\propto a
\Rightarrow t^{3}\propto a^{3}$$$$\rightarrow(according~
Eq.5)\rho_{q}\propto a^{-3}$$ On the other hand, because
$\rho_{m}\propto a^{-3}$ has $\Omega_{q}\simeq const$ which is in
conflict with the previous $\Omega_{q} \propto a^3$ and as a result
the agegraphic dark energy does not dominate and this is
unacceptable.
Two solutions have been proposed to solve this confusion[38]:\\
$Solution~ 1$: Placing $T+\delta$ instead of $T$ ($\delta$ is a
constant with time dimension)in results obtained in Equation $(5)$
[58]:
\begin{equation} \rho_{q}=\frac{3n^{2}m_{p}^{2}}{(T+\delta)^{2}} \end{equation}
In this solution, $T\ll\delta$ is at the early time, and as a result
$\rho_{q}\simeq const$, which means that the agegraphic dark energy
behaves like a cosmic constant. At the late time, $T\gg\delta$,in
this state $\rho_{q}$ is approximately equal with Equation $(5)$.
And in the intermediate state,$T\sim\delta$, in this state the
existence of $\delta$ can not be ignored and as a result there will
be no more tracking behavior. This solution overrules the basis of
the relationship in this model (karolyhazy relation $(1)$) and
becomes only a phenomenological model [38].\\
$Solution~ 2$: Because Equation $(5)$ is derived from Minkowski
space-time and on the other hand because space-time is very curved
in the early time, so this equation has no validity in the early
time and therefore $n=n(t)$ is variable [58]. In this solution, if
we consider a critical $T_c$ in the early or middle period,the
coefficient $n$ can be considered almost a constant in Equation (5),
thus with this method, confusion can be eliminated, although the
validity of the equation $(5)$ remove in early time[38].
\subsection{New agegraphic dark energy(NADE)}
In this section, we review a new model of agegraphic which is a
better model to solve the confusion mentioned above. According to
the vacuum energy density of the holographic
principle,$\rho_{\Lambda}=3c^{2}m_{p}^{2}L^{-2}$, If we consider $L$
as $\frac{1}{H}$, the equation of state parameter (EoS) of the
holographic dark energy is zero and therefore can not accelerate the
expansion of the universe [44]. Now, if we consider $L$ as the
particles horizon, in this case, because the parameter of the
equation of state becomes $\omega_{\Lambda}>\frac{-1}{3}$, this case
can not also describe the acceleration of the expansion of the
universe [59]. Finally, if $L$ be the the future event horizon of
the universe, in this case, the EoS holographic energy is equal to:
[44,59-61]:
\begin{equation}
\omega_{\Lambda}=\frac{-1}{3}-\frac{2\sqrt{\Omega_{\Lambda}}}{3c}
\end{equation}
Therefore, it is obvious that $\omega_{\Lambda}<\frac{-1}{3}$ and
can be a necessary condition for accelerating the expansion of the
universe. This model is called the holographic model, and because of
the future event horizon, it shows the eternal acceleration of the
universe but faces the causality problem. In the new dark energy
agegraphic model (NADE),instead of $T$ as the age of the universe,a
conformal time $\eta$ is chosen,defined as $d\eta=\frac{dt}{a}$ ($t$
is a cosmic time) to solve the causality problem [38]. Therefore,
the energy density of the NADE model is presented as:
\begin{equation} \rho_{q}=\frac{3n^{2}m_{p}^{2}}{\eta^{2}} \end{equation}
\begin{equation} \eta=\int\frac{dt}{a}=\int\frac{da}{a^2H} \end{equation}
and for fractional energy density
\begin{equation} \Omega_{q}=\frac{n^{2}}{H^{2}\eta^{2}} \end{equation}
For a flat FRW universe containing new agegraphic dark energy and
pressureless matter,using the energy conservation equation
$\dot{\rho}_{m}+3H\rho_{m}=0$ and equations $(6)$,$(13)$ and $(15)$
we have :[38]
\begin{equation}
\frac{d\Omega_{q}}{da}=\frac{1}{a}\Omega_{q}(1-\Omega_{q})(3-\frac{2\sqrt{\Omega_{q}}}{na})
\end{equation}
Also,from the energy conservation equation
$$\dot{\rho}_{q}+3H(\rho_{q}+p_{q})=0$$ and equations $(13)$ and
$(15)$, the equation of state (EoS) can be obtained for the NADE
model:
\begin{equation}\omega_{q}=-1+\frac{2\sqrt{\Omega_{q}}}{3na} \end{equation}
In late time, if $a\rightarrow\infty$ and $\Omega_q\rightarrow1$
then $\omega_{q}\rightarrow-1$ and in early time if $a\rightarrow0$
and $\Omega_q\rightarrow0$ then $\omega_q$ of relation $(17)$ Not
available. In the matter-dominated epoch we have:
$$H^{2}\propto\rho_{m}\propto a^{-3}\Rightarrow \sqrt{a}da\propto\frac{dt}{a}\propto d\eta$$$$\rightarrow(accoriding~ Eq.13~ and ~ \rho_{q}\propto\frac{1}{a}) \eta\propto\sqrt{a}$$
And from the energy conservation equation
$\dot{\rho}_{q}+3H\rho_{q}(1+\omega_{q})=0$,the parameter of the
state equation is equal to: $\omega_{q} =-\frac{2}{3}$.According to
$\rho_{m}\propto a^{-3}$,it follows that $\Omega_{q}\propto a^{2}$
and with comparing $\omega_{q}=-\frac{2}{3}$ and Equation $(17)$ :
\begin{equation} \Omega_{q}=\frac{n^{2}a^{2}}{4} \end{equation}
Thus the equation of motion in the matter-dominated epoch is
obtained:
\begin{equation} \frac{d\Omega_{q}}{da}=\frac{1}{a}\Omega_{q}(3-\frac{2\sqrt{\Omega_{q}}}{na})
\end{equation}
Now consider a radiation-dominated universe, which contains NADE
energy and background matter with the equation parameter
$\omega_{m}$ (for a particular case,$\omega_{m}=0$ for pressureless
matter and $\omega_{m} =\frac{1}{3}$ for radiation). Again,using the
energy conservation equation
$\dot{\rho}_{m}+3H\rho_{m}(1+\omega_{m})=0$ and equations $(6)$,
$(13)$ and $(15)$, the equation of motion for $\Omega_{q}$ is:[38]
\begin{equation}
\frac{d\Omega_{q}}{da}=\frac{1}{a}\Omega_{q}(1-\Omega_{q})(3(1+\omega_{m})-\frac{2\sqrt{\Omega_{q}}}{na})
\end{equation}
also in radiation-dominant epoch:
$$H^{2}\propto \rho_{r}\propto a^{-4}\Rightarrow ada\propto\frac{dt}{a}\propto d\eta$$$$\Rightarrow( according~ Eq.13~ and~  \rho_{q}\propto a^{-2}) \eta\propto a$$
And from the energy conservation equation
$\dot{\rho}_{q}+3H\rho_{q}(1+\omega_{q})=0$,the parameter of the
equation of state is equal to: $\omega_{q}=-\frac{1}{3}$ and
according to $\rho_{r}\propto a^{-4}$ it follows that
$\Omega_{q}\propto a^{2}$ and with comparing
$\omega_{q}=-\frac{1}{3}$ and Equation $(17)$:
\begin{equation} \Omega_{q}=n^{2}a^{2} \end{equation}
Thus the equation of motion in the radiation-dominated epoch is
obtained:[38]
\begin{equation} \frac{d\Omega_{q}}{da}=\frac{1}{a}\Omega_{q}(4-\frac{2\sqrt{\Omega_{q}}}{na})
\end{equation}
In summary, in the matter-dominated epoch $\omega_{q} =-\frac{2}{3}$
and $\Omega_{q} =\frac{n^{2}a^{2}}{4}$ and in the radiation-dominant
epoch, $\omega_{q}= -\frac{1}{3}$ and $\Omega_{q}= n^{2}a^{2}$,
which leads to domination of the new agegraphic dark energy. At the
late time $\omega_{q}\rightarrow -1$  and $a\rightarrow\infty$ that
NADE mimics a cosmic constant. Note that both the matter-dominated
epoch and the radiation-dominated epoch are be $\Omega_{q}\ll 1$ and
$a\ll 1$ [38]. Therefore, the NADE model is very different from the
ADE model and the evolutionary behavior of the NADE model is similar
to the holographic dark energy, except that the NADE model does not
have the causal problem [38]. In the NADE model, Equation $(1)$
(Karolyhazy equation) naturally follows
the entropy boundary of the holographic black hole [41].\\
\subsection{Interactive agegraphic dark energy(IADE)}
In this model, Wei [47] extended the NADE model by including the
interaction between the new agegraphic dark energy and the
background matter whose the equation of state parameter is
$\omega_{m} = const$. By including an interactive $Q$ expression,
Wei obtained the following relationships between the new agegraphic
dark energy and the background matter:
\begin{equation} \dot{\rho}_{q}+3H\rho_{q}(1+\omega_{q})=Q
\end{equation}
\begin{equation} \dot{\rho}_{m}+3H\rho_{m}(1+\omega_{m})=-Q
\end{equation}
The energy conservation equation in this case is $\dot{\rho}_{tot}+
3H(\rho_{tot} + p_{tot})=0$. Using equations $(6)$, $(13)$, $(15)$,
and $(24)$ the equation of motion for $\Omega_{q}$ in this model is
equal to:
\begin{equation}
\frac{d\Omega_{q}}{da}=\frac{1}{a}\Omega_{q}[(1-\Omega_{q})(3(1+\omega_{m})-\frac{2\sqrt{\Omega_{q}}}{na})-\frac{Q}{3m_{p}^{2}H^{3}}]
\end{equation}
Also from equations $(13)$,$(15)$ and $(23)$ the equation of state
parameter of the new agegraphic dark energy  is obtained as:[47,62]
\begin{equation} \omega_{q}= -1+
\frac{2\sqrt{\Omega_{q}}}{3na}-\frac{Q}{3H\rho_{q}} \end{equation}
According to Equation $(17)$, $\omega_{q}> -1$ and can not cross the
phantom divided line, if $Q=0$ (non- interaction mode) and return to
the NADE model again. But in the case $Q\neq0$ (interaction mode)
according to Equation $(25)$ if $Q<0$ then $\omega_{q}>-1$ and still
can not cross the phantom divided line, but if $Q>0$ then
$\omega_{q}>-1$ or $\omega_{q}<-1$ and therefore is likely to cross
the phantom divided line and it has a phantom behavior. In general,
the NADE model is the third single-parameter model after the
$\Lambda CDM$ model and the DGP braneworld model [63]. It is also
found in [63] that the coincidence problem in the NADE model is
naturally solvable and that this model is consistent with the cosmic
observations of Type Ia supernovae,the cosmic microwave
background,and the large scale structure. So the NADE model is a
very good and logical model for describing dark energy in cosmology.
The original dark energy agegraphic model (ADE) itself is limited by
the Big Bang nucleosynthesis (BBN) [64]. To reduce this limitation,
a coupling between dark energy and dark matter can be introduced
[65]. In addition, both the NADE model and the holographic dark
energy model have the problem of instabilities [48,49]. However, one
of the best models for describing dark energy is the NADE model, and
it can be a window to better understand the nature of dark
energy in the universe.\\
\section{Investigation of NADE model based on three types of GUP ($KMM$,$Nouicer$,$GUP^*$)}
In this section, we presents a new Agegraphic dark energy model
(NADE) based on the three types of generalized uncertainty
principles mentioned in [56]($KMM$, $Nouicer$ and $GUP^*$). This
model based on $KMM$ GUP has done before, but based on $Nouicer$ and
$GUP^*$ GUP is investigated for the first time in this paper.
\subsection{NADE model based on $KMM$ GUP:}
This part considers the KMM generalized uncertainty principle first
proposed by  Kim et al.[66] for the new agegraphic dark energy model
(NADE):
\begin{equation}
\Delta E \Delta t\geq 1+\beta(\Delta E)^2
\end{equation}
In unit $c=\hbar=k_{B} =1$. $\beta$ here, like the Planck length, is
a length scale $(\beta\sim \frac{1}{m_{p}}\sim L_{p})$. With the
solution of Eq.$(27)$, we reach $\Delta E_{G}=\frac{1}{\Delta t}+
\frac{\beta}{\Delta t^3}$(According to cosmological purpose such as
$\Delta t\sim t$, so it can also be written as $\Delta
E_{G}=\frac{1}{t}+\frac{\beta}{t^3}$). The energy density obtained
from GUP in this case is equal to [66]:
\begin{equation}
\rho_{G}=\frac{\Delta E_{G}}{\delta t^3}
\end{equation}
According to Karolyhazy time fluctuations, $\delta t=
t_{p}^\frac{2}{3}t^\frac{1}{3}$ results
$\beta=\frac{q^2}{n^2m_{p}^2}$, $t_{p}=\frac{1}{3n^2m_{p}^2}$ where
$q$ and $n$ are two constant parameters and $m_{p}$ is Planck's
mass. Energy density is defined by the parameters $q$ and $n$ and
Planck mass $m_{p}$ as follows:
\begin{equation}
\rho_{G}=\frac{3n^2m_{p}^2}{t^2}+\frac{3q^2}{t^4}
\end{equation}
Now, according to the new agegraphic dark energy model, instead of
the cosmic time parameter $t$, the age of the universe $T$, and then
to solve the causality problem in this model. Now, instead of $T$,it
is placed the conformal time $\eta$,which is defined as follows:
\begin{equation}
\eta=\int_{0}^{a}{\frac{da}{a^2H}}=\int_{-\infty}^{t}{\frac{dt}{aH}}
; \{t=lna ; H=\frac{\dot{a}}{a} \}
\end{equation}
Finally
\begin{equation}
\rho_{G}=\frac{3n^2m_{p}^2}{\eta^2} + \frac{3q^2}{\eta^4}
\end{equation}
On the other hand, Friedman's equations for a flat universe and also
its continuity equations are:
\begin{equation}
H^2=\frac{1}{3m_{p}^2}(\rho_{G}+\rho_{m});
\end{equation}
\begin{equation}
\dot{\rho_{G}}+3H(\rho_{G}+P_{G})=0
\end{equation}
\begin{equation}
\dot{\rho_{m}}+3H\rho_{m}=0;
\end{equation}
where $\rho_{G}$ and $P_{G}$ are the energy density and pressure
obtained from GUP respectively and $\rho_{m}$ are the density of
cold dark matter (CDM) without pressure ($P_{m}=0$). Introducing the
density parameters $\Omega_{i}=\frac{\rho_{i}}{3H^2m_{p}^2}$ and the
energy density equation obtained in this part and $\Omega_{G}+
\Omega_{m}=1$ we have:
\begin{equation}
\Omega_{G}=\frac{n^2}{H^2\eta^2}(1+ \frac{q^2}{m_{p}^2n^2\eta^2})
\end{equation}
According to $Eq.34$ and for the pressure expression we have:
\begin{equation}
P_{G}=-\frac{1}{3}\frac{d\rho_{G}}{dt}-\rho_{G}
\end{equation}
that as consequence the parameter of state equation is as follows:
\begin{equation}
\omega_{G}=\frac{P_{G}}{\rho_{G}}=-1+\frac{2e^{-t}\sqrt{\Omega_{G}}}{3n}(\frac{m_{p}^2n^2\eta^2+2q^2}{m_{p}^2n^2\eta^2+q^2})^\frac{2}{3}
\end{equation}
Using the expression $\Omega_{G} + \Omega_{m}=1$ and $Eq.34$ and
$Eq35$, the evolution equation to define $\omega_{G}$ and conformal
time are described as follows [66]:
\begin{equation}
\frac{d\Omega_{G}}{dt}=3\omega_{G}\Omega_{G}(1-\Omega_{G})
\end{equation}
\begin{equation}
\frac{d\eta}{dt}=\frac{1}{H_{0}}(\frac{H_{0}e^{-t}}{H})=\frac{1}{H_{0}}\sqrt{e^{t}\frac{1-\Omega_{G}}{\Omega_{m_{0}}}}
\end{equation}
where $H_{0}$ is the present Hubble parameter. According to the
equation of the state parameter in this part, it is expected that
dark energy will dominate in the present and the future. In the far
past($t\rightarrow-\infty$ or $a\rightarrow0$) the matter-dominated
universe with $\omega_{G}=-\frac{2}{3}$ and
$\Omega_{G}=\frac{n^2a^2}{4}$ and the radiation-dominated universe
with $\omega_{G}=-\frac{1}{3}$ and $\Omega_{G}=n^2a^2$ is recovered
[38]. According to [66], the evolution of the new agegraphic dark
energy model for the critical parameter $n$ and $q=1$ with the
initial conditions $\Omega_{G_{0}}=0.72$ and
$\eta_{0}=\frac{1}{H_{0}}$ has been investigated and shows that in
the far past, the matter-dominated ($\Omega_{G}\rightarrow0$ and
$\Omega_{m}\rightarrow1$)and in the far future, radiation-dominated
($\Omega_{G}\rightarrow1$ and $\Omega_{m}\rightarrow0$) will occur.
In other words, in terms of critical $n$ ($n_{c}=2.7999$) we have:

$$\{n<n_{c} ; \omega_{G}\rightarrow\infty\}$$
$$\{n=n_{c} ; \omega_{G}\rightarrow-\frac{2}{3}\}$$
$$\{n>n_{c} ; \omega_{G}\rightarrow-1\}$$
\begin{equation}
\end{equation}
Because of the GUP represents the Planck scale ($L = L_{p}$) and is
related to the Planck period ($t =t_{p}=10^{-43}s$), it describes
the reason of the cosmological constant problem well, and it does
not change  the dark energy-dominated universe in the present and
the
far future significantly[66].\\
\subsection{NADE model based on Nouicer GUP}
Nouicer's principle of generalized uncertainty is about field theory
in non-anticommutative superspace, which is defined in terms of
energy-time as follows[56]:
\begin{equation}
\Delta E\Delta t\geq e^{\beta\Delta E^2}
\end{equation}
So that in unit $c=\hbar=k_B=1$ it has the following solution:
\begin{equation}
\Delta E_{G}=\frac{\beta}{\Delta t^2ln\Delta t}\rightarrow(\Delta
t\sim t) \frac{\beta}{t^2lnt}
\end{equation}
According to the relation of the energy density obtained from GUP,
($\rho_{G}=\frac{\Delta E_{G}}{\delta t^3}$) and also $\delta
t=t_{p}^{\frac{2}{3}}t^{\frac{1}{3}}$,and
$t_{p}=\frac{1}{3n^2m_{p}^2}$ and $\beta=\frac{q^2}{n^2m_{p}^2}$,
the energy density in this case, in terms of parameters $q$ ,$n$ and
Planck mass,$m_{p}$, is defined as follows:
\begin{equation}
\rho_{G}=\frac{9n^2q^2m_{p}^2}{t^3lnt}
\end{equation}
Now for investigating Nouicer GUP for the NADE model in $Eq.43$
instead of $t$, we place cosmic age, $T$,and then to solve the
causality problem in this model instead of $T$,we place the
conformal time $\eta$ which we will finally have:
\begin{equation}
\rho_{G}=\frac{9n^2q^2m_{p}^2}{\eta^2 ln\eta}
\end{equation}
According to $Eq.32$,$Eq.33$,$Eq.34$, and the derivative of the
expression $\Omega_{G}+\Omega_{m}=1$ and the density parameter
$\Omega_{i}=\frac{\rho_{i}}{3H^2m_{p}^2}$ we have:
\begin{equation}
\Omega_{G}=\frac{3q^2n^2}{H^2\eta^3 ln\eta}
\end{equation}
Now using $Eq.36$ and $Eq.44$, the parameter of state equation for
this model is obtained as follows:
\begin{equation}
\omega_{G}=\frac{P_{G}}{\rho_{G}}=\frac{1}{3ln\eta}
\end{equation}
The equation of evolution related to $Eq.46$ is defined according to
the derivative of the expression $\Omega_{G} + \Omega_{m}=1$ and
also $Eq.34$ and $Eq.35$ and the conformal time as follows:
\begin{equation}
\frac{d\Omega_{G}}{dt}=-3\Omega_{G}(1+\omega_{G})
\end{equation}

\begin{figure}[h]
\begin{center}\includegraphics{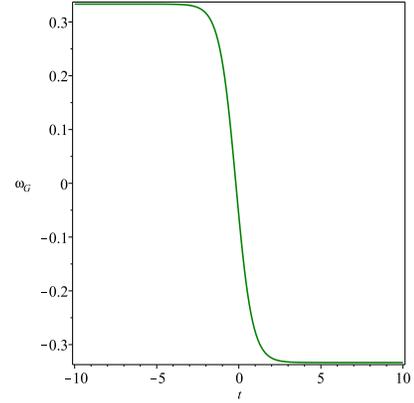} \vspace{12cm}\includegraphics{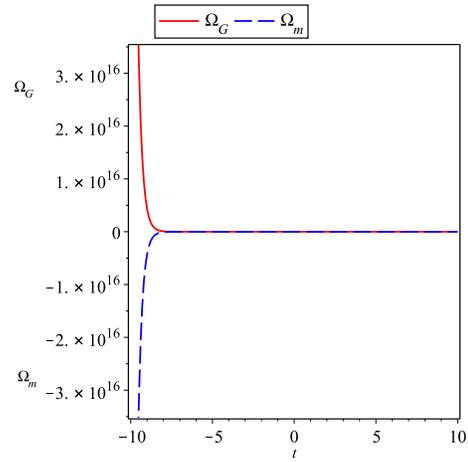}
\end{center}
\caption{\small {{\bf (Up).}Evolution of state parameter equation in
terms of the cosmic time in the Nouicer GUP. {\bf (Down).} Evolution
of density parameters of dark energy $\Omega_{G}$ and cold dark
matter $\Omega_{m}$ in the Nouicer GUP. With values $n=1.2$, $q=1$,
$m_{p}=1$, $H=-1$.(the values of the vertical axis are obtained
according to the values of the mentioned parameters.)}}
\end{figure}

As we see in $Fig.1.a$ and $Fig.1.b$, the behavior of the equation
of state parameter and the evolution equations for dark energy
($\Omega_{G}$) and cold dark matter ($\Omega_m$)are plotted in the
NADE model based on the Nouicer GUP. In $Fig.1.b$, according to the
Nouicer model,in the far past time,both matter and radiation are
dominated and tend to infinity ($\Omega_{G}\rightarrow+\infty$ and
$\Omega_{m}\rightarrow-\infty$)and have a repulsive effect from each
other, which causes accelerated expansion of the universe. With the
evolution of the universe and the passing of time, the effects of
$\Omega_{G}$ and $\Omega_{m}$ converge and neutralize each other's
effect, which means that $\Omega_{G}$ and $\Omega_{m}$ have no
effect on the evolution and expansion of the universe in the
direction of moving towards the present. But,with the effect of
$\omega_{G}$ and its decreasing behavior, the expansion of the
universe increase with the evolution of the universe. The
interesting note is that the higher the value of the parameter $n$,
especially when $n>n_{c}$, the universe expands more rapidly. In the
NADE model based on the Nouicer GUP because $\omega_{G}>-1$
therefore has a
quintessence-like behavior.\\
\subsection{NADE model based on $GUP^*$( The higher order  generalized uncertainty principle)}
The GUP is perturbative, meaning that for smaller values the GUP
parameter is set, and on the other hand the maximum momentum defined
in the general relativity is not set in GUP. To solve these
problems, the  higher order generalized uncertainty principle was
defined [56]. The energy-time uncertainty relationship of $GUP^*$ is
defined as follows:
\begin{equation}
\Delta E\Delta t\geq\frac{\hbar}{2(1-\beta\Delta E^2)}
\end{equation}
The solution of this equation is as follows (in unit
$c=\hbar=k_{B}=1$):
\begin{equation}
\Delta E_{G}=\frac{\hbar t}{2(t^2-\beta)}
\end{equation}
where $\beta$ is the GUP parameter and $\hbar$ is the reduced Planck
constant,
%which is defined as follows ($h$ is Planck constant
%and$h\simeq6.7×10 ^{-34}Js$):
\begin{equation}
\hbar=\frac{h}{2\pi}=\frac{6.7\times10^{-34}}{2\pi}\simeq1.06633\times10^{-34}Js
\end{equation}
According to the relation of the energy density obtained from GUP,
($\rho_{G}=\Delta E_{G}/\delta t^3$) and also $\delta
t=t_{p}^{\frac{2}{3}}t^{\frac{1}{3}}$,and
$t_{p}=\frac{1}{3n^2m_{p}^2}$ and $\beta=\frac{q^2}{n^2m_{p}^2}$,
 the energy density in this case,in terms of parameters $q$ and $n$
and Planck mass,$m_{p}$ is defined as follows:
\begin{equation}
\rho_{G}=\frac{9n^6m_{p}^6\hbar}{2(t^2n^2m_{p}^2-q^2)}
\end{equation}
For investigating $GUP^*$ in the NADE model in $Eq.43$ instead of
$t$, we place cosmic age, $T$, and then to solve the causality
problem in this model instead of $T$, we place the conformal time
$\eta$ which we will finally have:
\begin{equation}
\rho_{G}=\frac{9n^6m_{p}^6\hbar}{2(\eta^2n^2m_{p}^2-q^2)}
\end{equation}
According to $Eq.32$,$Eq.33$,$Eq.34$, and the derivative of the
expression $\Omega_{G}+\Omega_{m}=1$ and the density parameter
$\Omega_{i}=\frac{\rho_{i}}{3H^2m_{p}^2}$ we have:
\begin{equation}
\Omega_{G}=\frac{3\hbar m_{p}^4n^6}{2H^2(\eta^2m_{p}^2n^2-q^2)}
\end{equation}
Now according to $Eq.36$ and $Eq.52$,the parameter of state equation
for this model is obtained as follows:
\begin{equation}
\omega_{G}=-1+\frac{2\eta^2}{3(\eta^2m_{p}^2n^2-q^2)}
\end{equation}
The equation of evolution related to $Eq.46$ is defined according to
the derivative of the expression $\Omega_{G}+\Omega_{m}=1$ and also
$Eq.34$ and $Eq.35$ and the conformal time as follows:
\begin{equation}
\frac{d\Omega_{G}}{dt}=3\Omega_{G}(\omega_{G}(\eta^2m_{p}^2n^2-q^2)-q)
\end{equation}

\begin{figure}[h]
\begin{center}\includegraphics{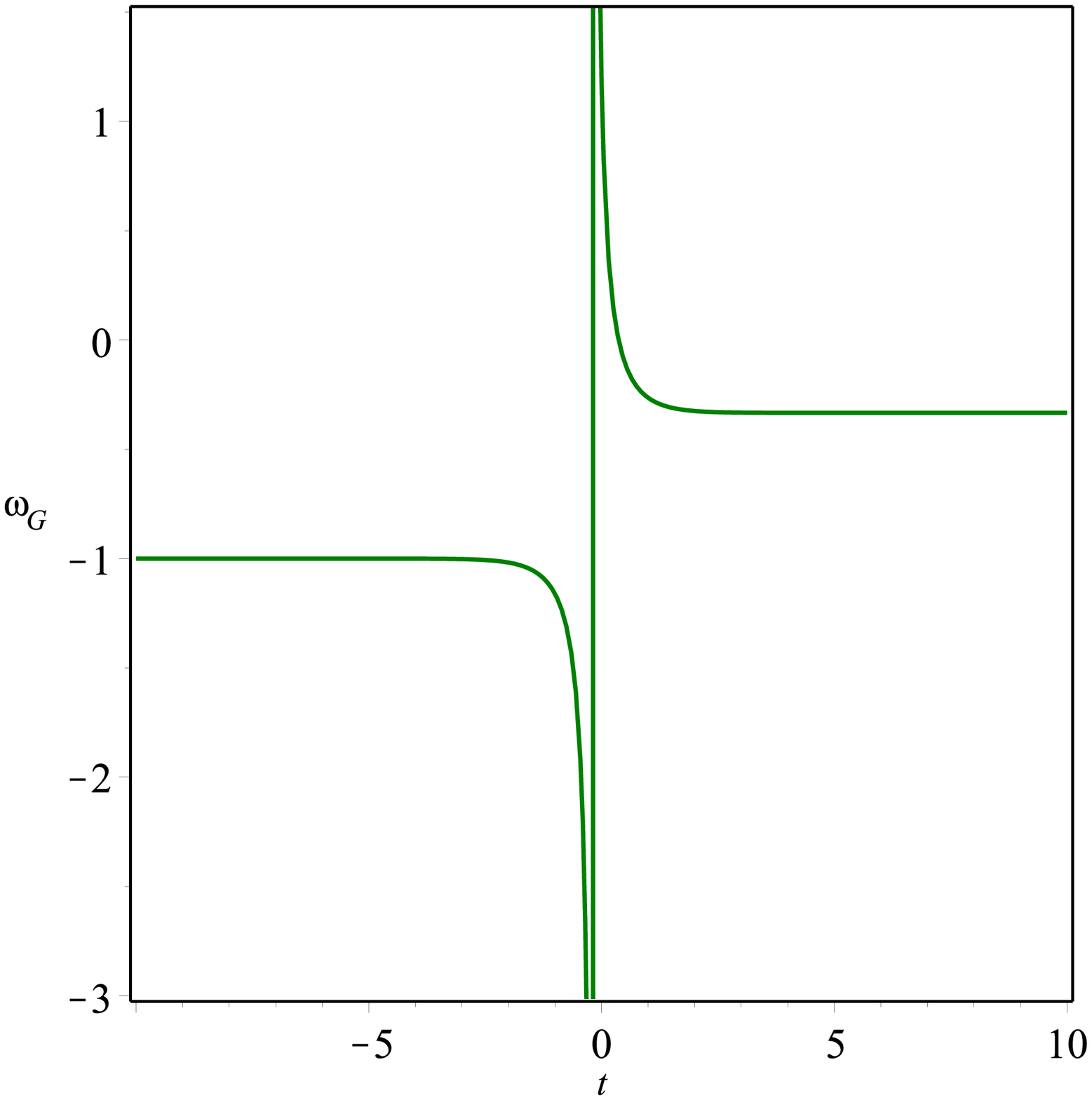} \vspace{12cm}\includegraphics{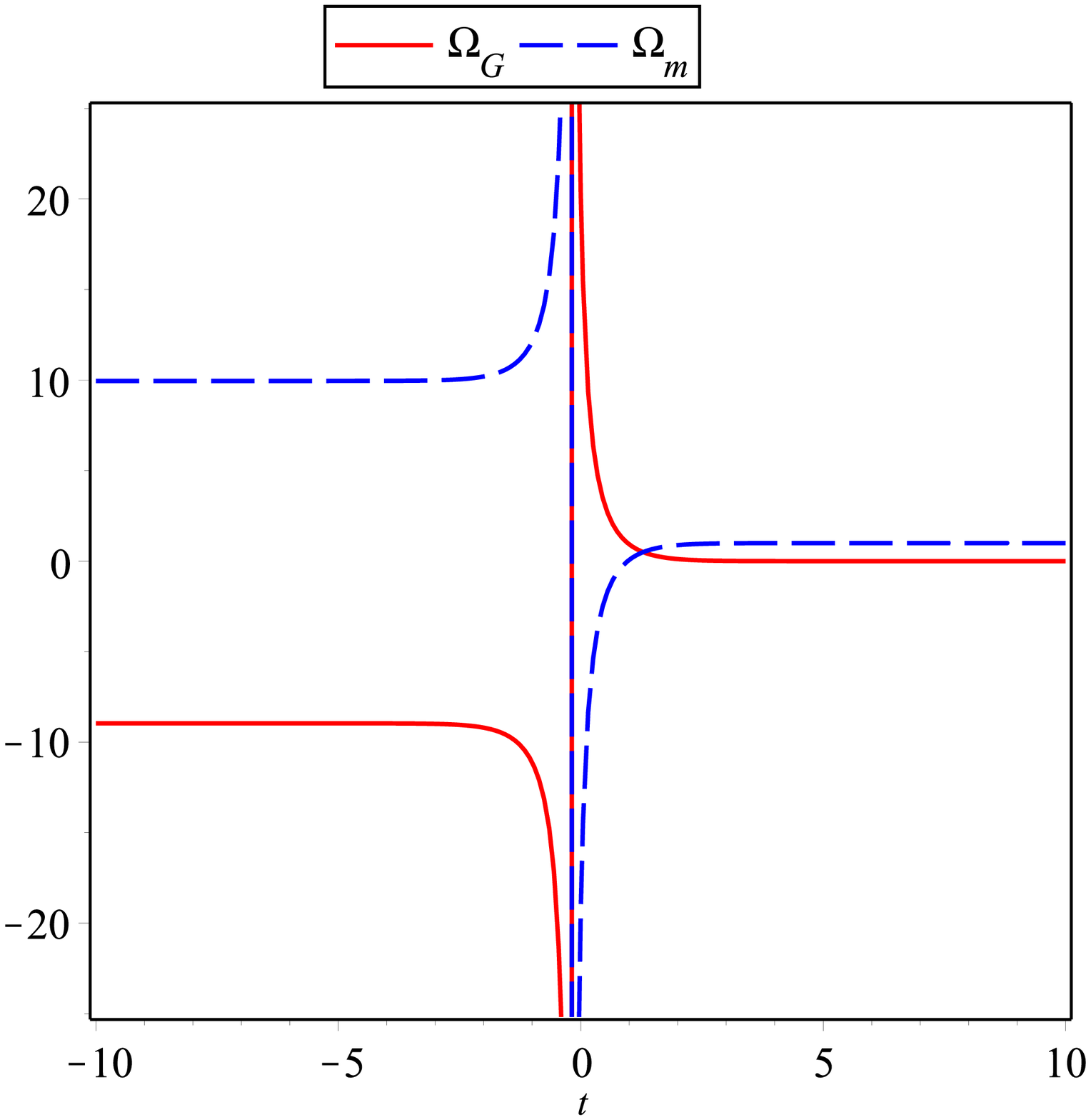}
\end{center}
\caption{\small {{\bf [a](Up).} Evolution of state parameter
equation in terms of the cosmic time in the $GUP^*$. {\bf
[b](Down).} Evolution of density parameters of dark energy
$\Omega_{G}$ and cold dark matter $\Omega_{m}$ in the $GUP^*$ . With
values $n=1.2$, $q=1$, $m_{p}=1$, $H=-1$,$\hbar=2$ (the values of
the vertical axis are obtained according to the values of the
mentioned parameters.)}}
\end{figure}

In $Fig.2.a$ and $Fig.2.b$, the behavior of the state parameter
equation and the evolution of dark energy $\Omega_{G}$ and cold dark
matter $\Omega_{m}$ are plotted in the NADE model based on $GUP^*$,
respectively. In $Fig.2.a$, in the far past time,  $\omega_{G}$ has
a decreasing behavior and because with the evolution of the
universe, the value of the state parameter is $\omega_{G}= -1$ and
then decreases rapidly ($\omega_{G}<-1$), so most of the time in the
far past, It has a same cosmological constant behavior and then
shows phantom-like behavior. And in the far future, $\omega_{G}$ has
a decreasing and quintessence-like behavior ($\omega_{G}>-1$) and at
the present time $\omega_{G}$ does not have a certain value. In
$Fig.2.b$, in the far past time, tend to
$\Omega_{G}\rightarrow-\infty$ and $\Omega_{m}\rightarrow+\infty$
that indicate in the far past, both matter and radiation is
dominated and have a repulsive effect on each other that it causes
the accelerated expansion of the universe. And in the far future
time $\Omega_{G}$ and $\Omega_{m}$ converge to a certain constant
value,and both in the far past and far future times all three
$\omega_{G}$,$\Omega_{G}$ and $\Omega_{m}$ are dominated. In the
$Table.1$,it summarizes the results obtained in this
section.\\
\begin{table}[p]
\begin{center}
\caption{The evolution of the universe in three types of GUP(KMM ,
Nouicer and GUP*).} \vspace{0.5 cm}
\begin{tabular}{|p{0.9cm}|p{1.9cm}|p{1.9cm}|p{1.9cm}|}
  \hline
  \hline {\tiny GUP} & {\tiny The far past time} & {\tiny The present time} & {\tiny The far future time}\\
  \hline {\tiny KMM}& {\tiny The matter dominated. \newline $\omega_G\rightarrow\infty$\newline$\Omega_m\rightarrow 1$\newline$\Omega_G\rightarrow 0$ }& {\tiny Increasing the radiation effect with the universe evolution.\newline $\omega_G\rightarrow\ -\frac{2}{3}$\newline$\Omega_m\rightarrow 0$\newline$\Omega_G\rightarrow 1$} & {\tiny The radiation dominated. \newline $\omega_G\rightarrow\ -1$\newline$\Omega_m\rightarrow 0$\newline$\Omega_G\rightarrow 1$}\\
  \hline {\tiny Nouicer}& {\tiny Both the matter and radiation dominated.\newline $\omega_G\rightarrow 1$\newline$\Omega_m\rightarrow -\infty$\newline$\Omega_G\rightarrow \infty$ }& {\tiny Decreasing the matter and radiation effect and increasing the effect of state parameter equation.}& {\tiny The state parameter equation dominated. \newline $\omega_G\rightarrow -1$\newline$\Omega_m\rightarrow 0$\newline$\Omega_G\rightarrow 0$}\\
  \hline {\tiny GUP*} &{\tiny The matter and radiation and the parameter of state equation dominated.\newline $\omega_G\rightarrow -\infty$\newline$\Omega_m\rightarrow \infty$\newline$\Omega_G\rightarrow -\infty$ } &{\tiny Lack of the matter and radiation and the parameterstate of state equation dominated.}& {\tiny The matter and
radiation and the parameter of state equation dominated and faster expansion rate than far past time.\newline $\omega_G\rightarrow  -0.5 $\newline$\Omega_m\rightarrow 1$\newline$\Omega_G\rightarrow  -1$}\\
  \hline
\end{tabular}
\end{center}
\end{table}

\section{Investigation of Emergent, Intermediate and Logamediate scenarios in GUP types (KMM, Nouicer, $GUP^*$)}
In this section, we examine the three scenarios mentioned for the
universe in[67,68] in the types of GUPs mentioned in this paper, and
check the evolution of the universe in each case.As respects the
three scenarios mentioned for the NADE model are based on the KMM
GUP in the various works previously reviewed (see;[20,46 and 66]),
for this reason we regardless from re-describing it ,only is
summarized in $Table.2$ at the end of this section,so in this
section we will concentrate on two types of GUP (Nouice and $GUP^*$)
with and without interaction.
\subsection{Emergent scenario based on Nouice GUP}
The scale factor in the emergent scenario is defined as follows[67]:
\begin{equation}
a(t)=a_{0}(B+e^{At})^m ; {a_{0}>0 , A>0 , B>0}
\end{equation}
Now if we place the above relation in the general expression of
$Eq.30$,we will have: \small
$$\eta=$$
\begin{equation}
-\frac{(1+Be^{-At})^m(B+e^{At})^{-m}\,_{2}F_{1}[m,m,1+m,-Be^{-At}]}{Aa_{0}m}
\end{equation}
\normalsize where $\,_{2}F_{1}$ is a hypergeometric function defined
as
$\,_{2}F_{1}=[a,b,c,z]=\sum_{i=0}^{\infty}\frac{\prod_{i=0}^{\infty}a
\prod_{i=0}^{\infty}b  z^{i}}{\prod_{i=0}^{\infty}c  i!}$ By placing
$Eq.57$ in $Eq.44$, the energy density in this part is obtained as
follows:\tiny
$$\rho_{G1}=$$
$$\Bigg({-9A^3a_{0}^3m^3(1+Be^{-At})^{-3m}(B+e^{At})^{3m}n^2q^2m_{p}^2}\Bigg)\times$$
$$\Bigg(\bigg(_{2}F_{1}[m,m,1+m,-Be^{-At}]^3\bigg)\times$$
\begin{equation}
\bigg(ln(-\frac{(1+Be^{-At})^m(B+e^{At})^{-m}\,_2F_{1}[m,m,1+m,-Be^{-At}]}{Aa_{0}m})\bigg)\Bigg)^{-1}
\end{equation}
\normalsize According to the conservation $Eq.24$ and $Eq.56$ in
this part we have for dark matter density:
\begin{equation}
\rho_{m1}=\rho_{m_{0}}[a_{0}(B+e^{At})^m]^{-3(1+\omega_{m}+Q)}
\end{equation}
where $Q$ is the expression of interaction between dark energy of
NADE model based on GUP with dark matter without pressure, which in
all parts,two modes of interaction ($Q\neq0$) and non-interaction($Q
=0$) are considered. Using $Eq.56$ in this part, the Hubble
parameter in terms of cosmic time is calculated as follows:
\begin{equation}
H=\frac{\dot{a}}{a}=\frac{mAe^{At}}{B+e^{At}}
\end{equation}
Using the conservation $Eq.23$ and $Eq.60$, in this part the dark
energy pressure is calculated as follows:
\begin{equation}
p_{G1}=\frac{-Q-\dot{\rho_{G1}}}{3H}-\rho_{G1}
\end{equation}
and using $Eq58$,$Eq.59$,$Eq.61$, the equation of the total state
parameter, $\omega_{total}$, is obtained as follows:
\begin{equation}
\omega_{total}=\frac{p_{G1}}{\rho_{G1}+\rho_{m1}}
\end{equation}
The behavior of the $\omega_{total}$ state parameter equation, given
in $Eq.62$,is plotted in $Fig.3$ in both the interaction and
non-interaction modes under the emergent scenario based on the
Nouicer GUP.

\begin{figure}[h]
\begin{center}\includegraphics{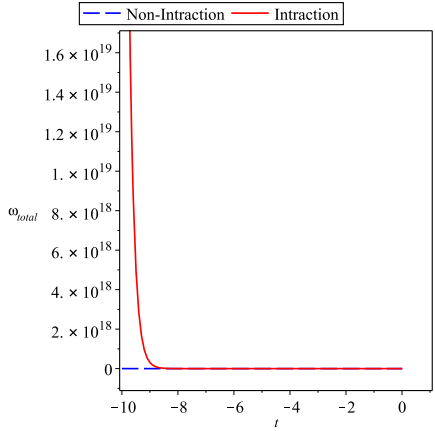} \vspace{6.5cm}
\end{center}
 \caption{\small {Evolution of state parameter equation in terms of
the cosmic time under the emergent scenario of the universe based on
the Nouicer GUP. With values $n=1.2$, $q=1$, $m_{p}=1$, $Q=0.05$,
$a_{0}=0.12$, $B=2.3$, $A=5.6$, $m=2$ (the values of the vertical
axis are obtained according to the values of the mentioned
parameters.)}}
\end{figure}

According to $Fig.3$, the emergent scenario based on the Nouicer GUP
is only able to describe the evolution of the universe in the far
past and present time, and can not describe the far future time. In
the mode of interaction with the evolution of the universe, the
value of $\omega_{total}$ decreases, and this indicates the
accelerated expansion of the universe in the far past, and also in
the mode of non-interaction in this part, the universe expands with
a constant rate in all of the past period. Because of the value of
$\omega_{total}>-1$, the universe behaves a quintessence-like in the
emergent scenario under the Nouicer GUP. In $Fig.4.a$ and $Fig.4.b$,
the parameters $\rho_{total}$ and $p$ are plotted in two modes of
interaction and non-interaction,respectively.
\begin{figure}[h]
\begin{center}\includegraphics{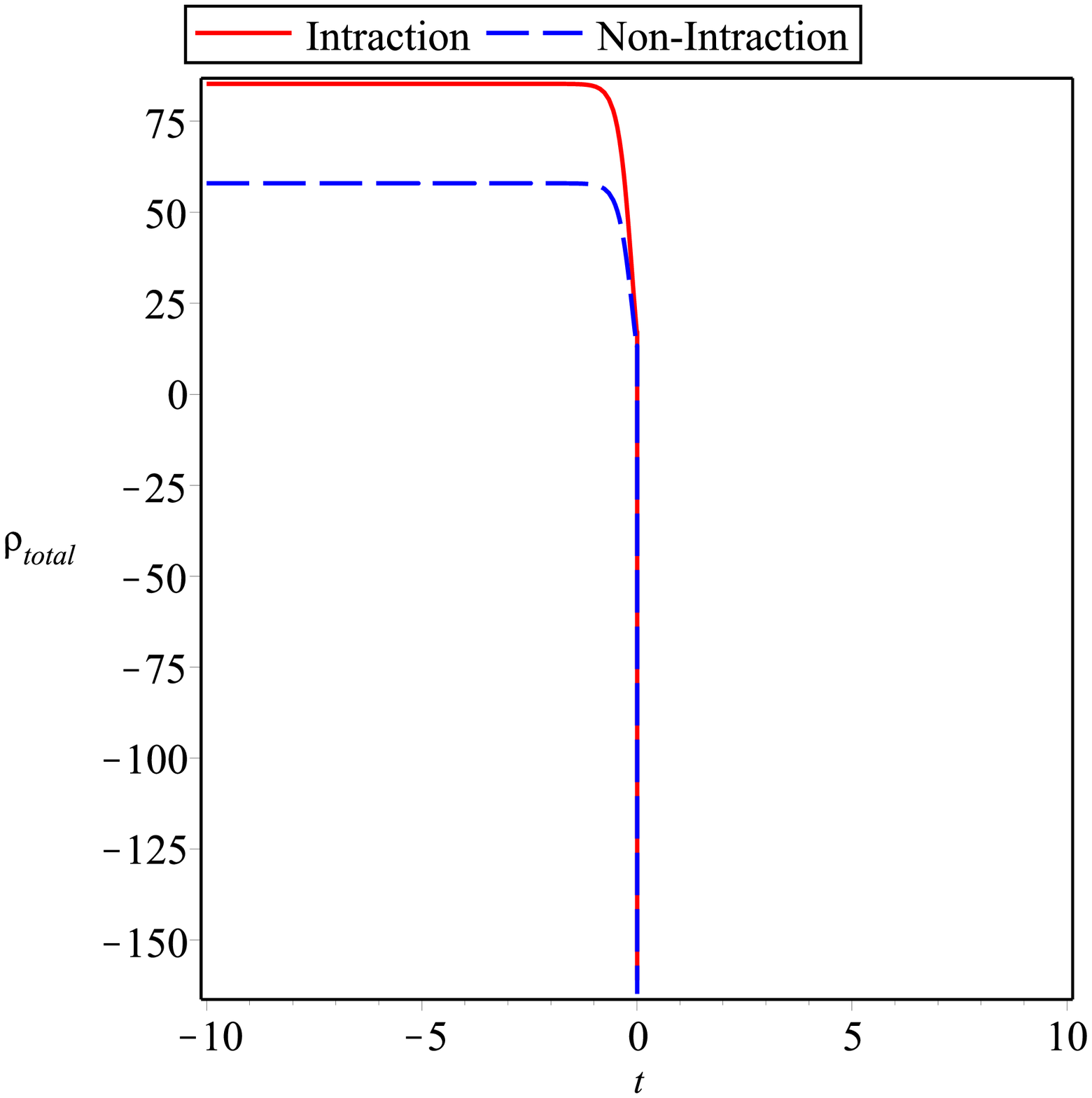} \vspace{12cm}\includegraphics{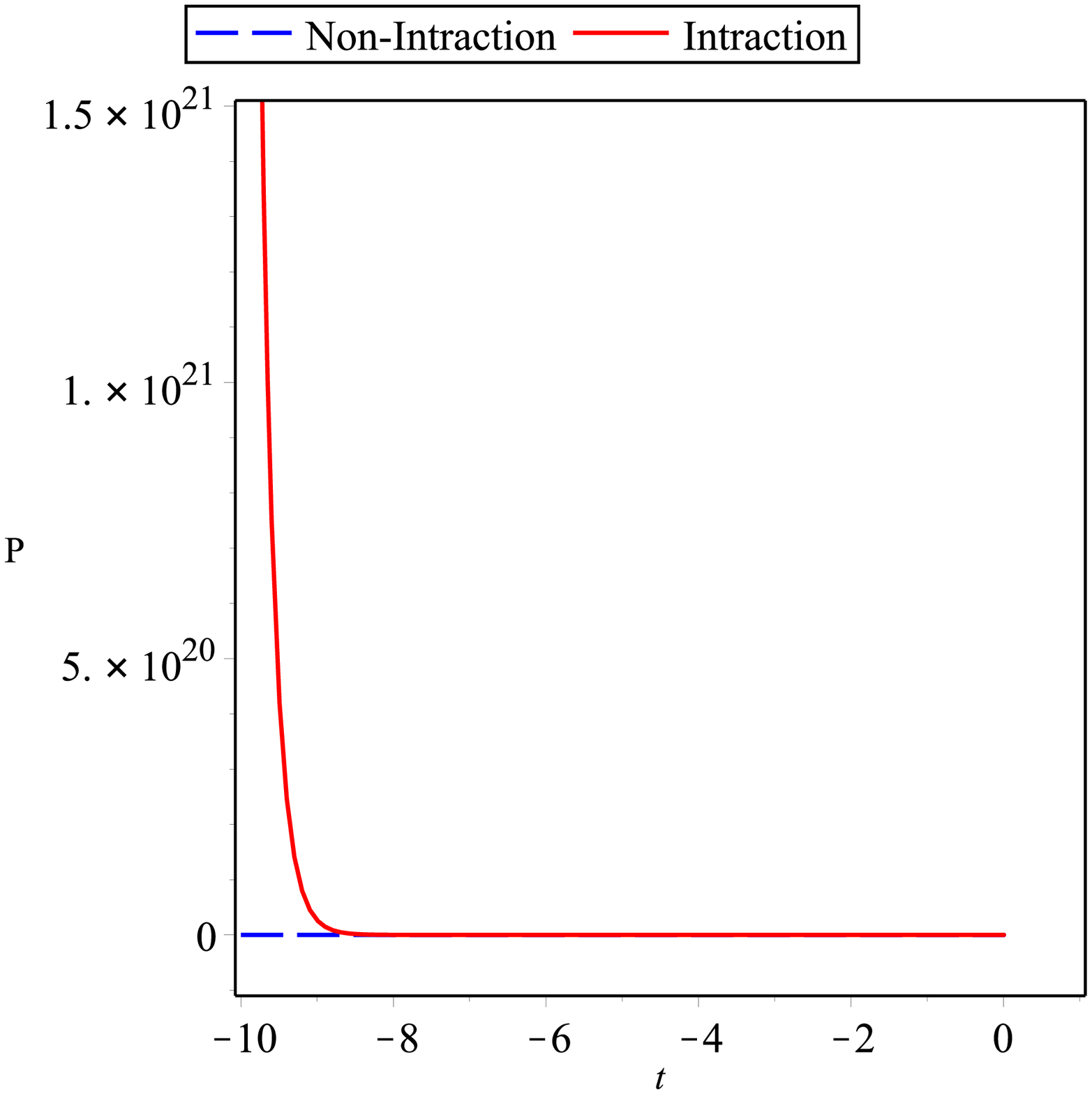}
\end{center}
\caption{\small {{\bf [a](Up).} Evolution of energy density
parameter,$\rho_{total}$,in terms of the cosmic time under the
emergent scenario of the universe based on the Nouicer GUP. {\bf
[b](Down).} Evolution of pressure parameter $P$ in terms of the
cosmic time under the emergent scenario of the universe based on the
Nouicer GUP .With values $n=1.2$, $q=1$, $m_{p}=1$, $Q=0.05$,
$a_{0}=0.12$, $B=2.3$, $A=5.6$, $m=2$ (the values of the vertical
axis are obtained according to the values of the mentioned
parameters.)}}
\end{figure}

In $Fig.4.a$, where $\rho_{total}$ values remain constant in both
interaction and non-interaction modes, then its value decreases
rapidly in a shorter time. And in $Fig.4.b$,the behavior of the $P$
parameter in both interaction and non-interaction modes is similar
to the behavior of the parameter of state equation in $Fig.3$, and
its reason is obviously due to $Eq.62$, and $\rho_{total}$ constant
remaining. To examine the stability or instability of different dark
energy models, a parameter called the speed square sound is
expressed, which Kim and Wei first proposed for their new agegraphic
model, which is defined as follows [49]:
\begin{equation}
\nu_{s}^2=\frac{\dot{P}}{\dot{\rho}}
\end{equation}
In $Fig.5$, in both interaction and non-interaction modes, the sound
speed square,$\nu_{s}^2$, are plotted under the Nouicer GUP emergent
scenario of the universe, which behaves similarly to the parameter
of state equation in this part. According to $Fig.5$,the behavior of
$\nu_{s}^2$ with the evolution of the universe in the far past has a
decreasing behavior and always remains at the non-negative level,
which indicates the stability of the NADE model under the emerging
scenario based on the Nouicer GUP.
\begin{figure}[h]
\begin{center}\includegraphics{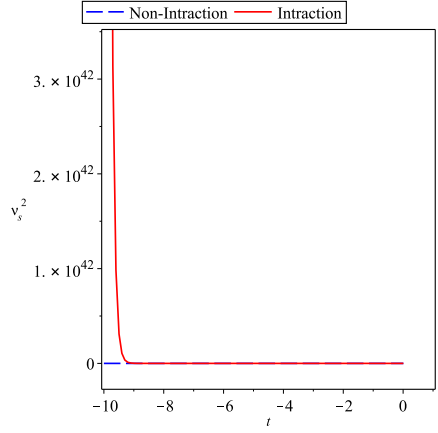} \vspace{6.5cm}
\end{center}
 \caption{\small {Evolution of the sound speed squares in terms of
the cosmic time to check the stability or instability of the dark
energy model under the emergent scenario of the universe based on
the Nouicer GUP. With values $n=1.2$, $q=1$,$m_{p}=1$, $Q=0.05$,
$a_{0}=0.12$, $B=2.3$, $A=5.6$, $m=2$ (the values of the vertical
axis are obtained according to the values of the mentioned
parameters.)}}
\end{figure}

\subsection{Intermediate scenario based on Nouicer GUP}
The scale factor in the emergent scenario is defined as follows
[68]:
\begin{equation}
a(t)=e^{\lambda t^{\beta}} ; \{\lambda>0 , 0<\beta<1 \}
\end{equation}
Now if we place the above relation in the general expression of
$Eq.30$, we will have:
\begin{equation}
\eta=-\frac{\lambda^{-\frac{1}{\beta}}\Gamma[\frac{1}{\beta},\lambda
t^{\beta}]}{\beta}
\end{equation}
where $\Gamma[x,z]$ is an incomplete gamma function defined as
$\Gamma[x,z]=\int_{z}^{\infty}t^{x-1}e^{-t}dt$. With placing $Eq.65$
in $Eq.44$, we have:
\begin{equation}
\rho_{G2}=-\frac{9n^2q^2m_{p}^2\beta^3\lambda^{\frac{3}{\beta}}}{\Gamma[\frac{1}{\beta},\lambda
t^{\beta}]^3ln(-\frac{\lambda^{-\frac{1}{\beta}}\Gamma[\frac{1}{\beta},\lambda
t^{\beta}]}{\beta})}
\end{equation}
According to the conservation $Eq.24$ under intermediate scenario we
have:
\begin{equation}
\rho_{m2}=\rho_{m_{0}}(e^{\lambda t^{\beta}})^{-3(1+\omega_{m}+Q)}
\end{equation}
where $Q$ is the expression of interaction between dark energy of
NADE model based on GUP with dark matter without pressure, which in
all parts,two modes of interaction($Q\neq0$) and
non-interaction($Q=0$) are considered. The Hubble parameter
according to $Eq.64$ in this part is:
\begin{equation}
H=\frac{\dot{a}}{a}=\lambda\beta t^{-1+\beta}
\end{equation}
According to the conservation $Eq.23$ and $Eq.68$,in this part the
dark energy pressure is calculated as follows:
\begin{equation}
p_{G2}=\frac{-Q-\dot{\rho_{G2}}}{3H}-\rho_{G2}
\end{equation}
The parameter of state equation and the square of the speed of sound
in this part according to the $Eq.66$,$Eq.67$ and $Eq.69$ are
respectively:
\begin{equation}
\omega_{total}=\frac{p_{G2}}{\rho_{G2}+\rho_{m2}}
\end{equation}
\begin{equation}
\nu_{s}^2=\frac{\dot{p_{G2}}}{\dot{\rho_{G2}}+\dot{\rho_{m2}}}
\end{equation}
$Fig.6$ shows the behavior of the $\omega_{total}$ state parameter
equation in $Eq.70$ given under the intermediate scenario based on
the Nouicer GUP. According to $Fig.6$,the intermediate scenario,
like the emergent scenario based on the Nouicer GUP, can only
describe the evolution of the universe in the far past time, and can
not describe the evolution of the universe in the present and the
far future time. In this part,the mode of interaction and
non-interaction have exactly the same behavior, in other
words,interaction has no effect on the evolution behavior of the
universe. As we see in $Fig.6$, with the evolution of the universe
$\omega_{total}$ has an increasing behavior, which means that in the
far past,unlike the emergent scenario in the previous part, in this
scenario, based on the Nouicer GUP, the universe does not have
accelerated expansion but has decelerated behavior. Because of in
this part is $\omega_{total}> -1$,so the universe has a
quintessence-like behavior.
\begin{figure}[h]
\begin{center}\includegraphics{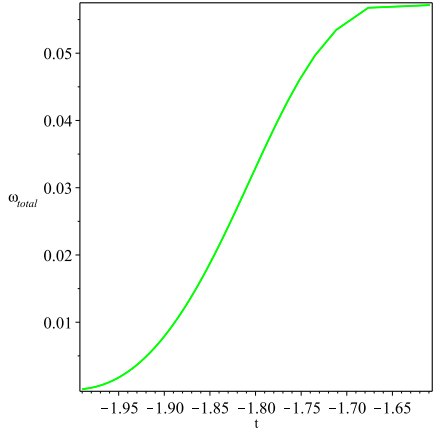} \vspace{6.5cm}
\end{center}
 \caption{\small {Evolution of state parameter equation in terms of
the cosmic time under the intermediate scenario of the universe
based on Nouicer GUP. With values $n=2$, $q=1$, $m_{p}=1$, $Q=0.05$,
$\lambda=1.9$, $\beta=0.5$ (the values of the vertical axis are
obtained according to the values of the mentioned parameters.)}}
\end{figure}

In $Fig.7.a$ and $Fig.7.b$, the parameters $\rho_{total}$ and $P$
are plotted in terms of cosmic time, respectively. In these
diagrams,the mode of interaction and non-interaction have exactly
the same behavior, and with the evolution of the universe, the
passing of time, the value of $\rho_{total}$ increases and the value
of $P$ increases in the negative direction($t\rightarrow-\infty$) in
other words with the evolution of the universe, the value of
negative $P$ is decreased.
\begin{figure}[h]
\begin{center}\includegraphics{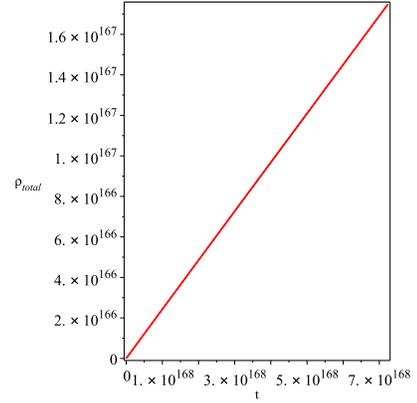} \vspace{12cm}\includegraphics{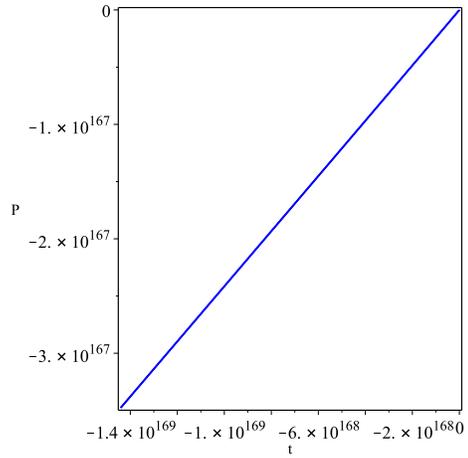}
\end{center}
\caption{\small {{\bf [a](Up).} Evolution of energy density
parameter,$\rho_{total}$, in terms of the cosmic time under the
intermediate scenario of the universe based on Nouicer GUP. {\bf
[b](Down).} Evolution of pressure parameter $P$ in terms of the
cosmic time under the intermediate scenario of the universe based on
Nouicer GUP . With values $n=2$, $q=1$, $m_{p}=1$, $Q=0.05$,
$\lambda=1.9$, $\beta=0.5$ (the values of the vertical axis are
obtained according to the values of the mentioned parameters.)}}
\end{figure}

$Fig.8$ shows the sound speed square behavior,$\nu_{s}^2$ , for the
stability or instability of the model in this part. As shown in
$Fig.8$, with the evolution of the universe and the passing of time,
$\nu_{s}^2$ has an increasing behavior in this part and always
remains at a positive level($\nu_{s}^2>0$), which indicates the
stability of the NADE model in the intermediate scenario based on
Nouicer GUP.

\begin{figure}[h]
\begin{center}\includegraphics{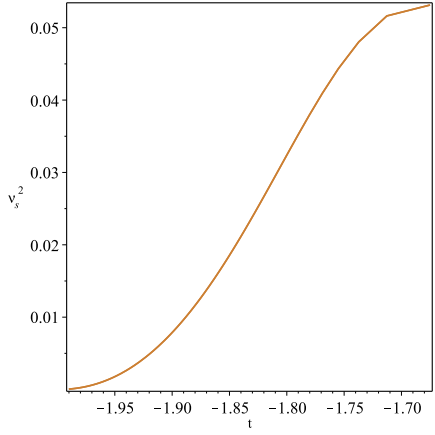} \vspace{6.5cm}
\end{center}
 \caption{\small {Evolution of the sound speed squares in terms of
the cosmic time to check the stability or instability of the dark
energy model under the intermediate scenario of the universe based
on Nouicer GUP. With values $n=2$, $q=1$, $m_{p}=1$, $Q=0.05$,
$\lambda=1.9$, $\beta=0.5$ (the values of the vertical axis are
obtained according to the values of the mentioned parameters.)}}
\end{figure}

\subsection{ Logamediate scenario based on Nouicer GUP}
The scale factor in the emergent scenario is defined as follows[68]:
\begin{equation}
a(t)=e^{\mu(lnt)^{\alpha}} ; \{ \alpha>1 , \mu>0 \}
\end{equation}
Now if we place the above relation in the general expression of
Equation 30, we will have:
\begin{equation}
\eta=\int\frac{dt}{e^{\mu(lnt)^{\alpha}}}
\end{equation}
Due to the lack of a analytic solution of the above conformal time,
and to solve this problem, we consider the above conformal time for
the specific case $\alpha=2$, which we will have:
\begin{equation}
\eta=\int\frac{dt}{e^{\mu}(lnt)^{2}}=\frac{e^{\frac{1}{4\mu}}\sqrt{\pi}erf[\frac{-1+2\mu
lnt}{2\sqrt{\mu}}]}{2\sqrt{\mu}}
\end{equation}
where $erf$ is an error function and is defined as
$erf=\frac{2}{\sqrt{\pi}}\int_{0}^{x}e^{-t^{2}}dt$. By placing
$Eq.74$ in $Eq.44$, the energy density in this part is obtained as
follows:
\begin{equation}
\rho_{G3}=\frac{72n^2q^2m_{p}^2\mu^{\frac{3}{2}}}{e^{\frac{3}{4\mu}}\pi^{\frac{3}{2}}erf(\frac{-1+2\mu
lnt}{2\sqrt{\mu}})^3ln(\frac{e^{\frac{1}{4\mu}}\sqrt{\pi}erf(\frac{-1+2\mu
lnt}{2\sqrt{\mu}})}{2\sqrt{\mu}})}
\end{equation}
According to the conservation $Eq.24$ and $Eq.72$ under the
logamediate scenario we have:
\begin{equation}
\rho_{m3}=\rho_{m_{0}}(e^{\mu(lnt)^{\alpha}})^{-3(1+\omega_{m}+Q)}
\end{equation}
where $Q$ is an expression related to interaction.  According to the
scale factor equation in the logamediate scenario ($Eq.72$),the
Hubble parameter in this part is calculated as follows:
\begin{equation}
H=\frac{\dot{a}}{a}=\frac{\mu\alpha(lnt)^{-1+\alpha}}{t}
\end{equation}
According to survival $Eq.24$ and $Eq.77$,in this part,the
expression pressure is obtained as follows:
\begin{equation}
p_{G3}=\frac{-Q-\dot{\rho_{G3}}}{3H}-\rho_{G3}
\end{equation}
The parameter of state equation and the square of the speed of sound
in this part, as in the previous parts, are as follows:
\begin{equation}
\omega_{total}=\frac{p_{G3}}{\rho_{G3}+\rho_{m3}}
\end{equation}
\begin{equation}
\nu_{s}^2=\frac{\dot{p_{G3}}}{\dot{\rho_{G3}}+\dot{\rho_{m3}}}
\end{equation}
$Fig.9$ shows the behavior of the state parameter equation in the
logamediate scenario based on the Noicer GUP. In this scenario,as in
the intermediate scenario, the interaction and non-interaction modes
behave exactly the same. Unlike the previous two scenarios, this
scenario is only able to describe the universe in the far future
time and is not able to describe the universe in the present and the
far past times. As shown in $Fig.9$,with the evolution of the
universe and the passing of time,$\omega_{total}$ has a decreasing
behavior, which means that the universe has an accelerated expansion
with the evolution and passing of time. Because of in this part is
$\omega_{total}>-1$ and in certain time is $\omega_{total}=-1$ ,so
the universe has a quintessence-like  and cosmological constant
-like behavior respectively.
\begin{figure}[h]
\begin{center}\includegraphics{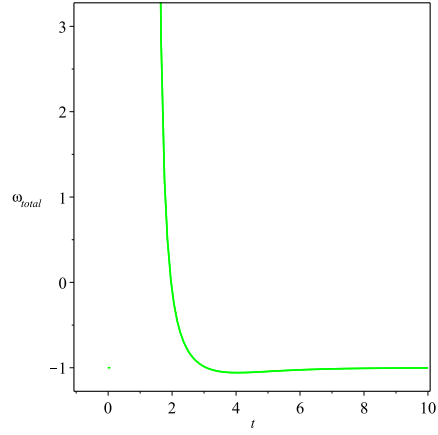} \vspace{6.5cm}
\end{center}
 \caption{\small {Evolution of state parameter equation in terms of
the cosmic time under the logamediate scenario of the universe based
on Nouicer GUP. With values $n=1.0013$, $q=1$, $m_{p}=1$, $Q=0.05$,
$\alpha=2$, $\mu=1.4$ (the values of the vertical axis are obtained
according to the values of the mentioned parameters.)}}
\end{figure}

In $Fig.10.a$ and $Fig.10.b$, the parameters $\rho_{total}$ and $P$
are plotted in terms of cosmic time,respectively. In these
diagrams,the two modes of interaction and non-interaction behave
exactly the same. with the evolution of the universe and the passing
of time,the value of $\rho_{total}$ increases rapidly and then
remains constant and always remains at a negative level
($\rho_{total}<0$). And in $Fig.10.b$,the value of $P$ also
increases rapidly and then remains constant,and after the value of
$\omega_{total}$ is constant,value of $P$ changes from negative to
positive level.
\begin{figure}[h]
\begin{center}\includegraphics{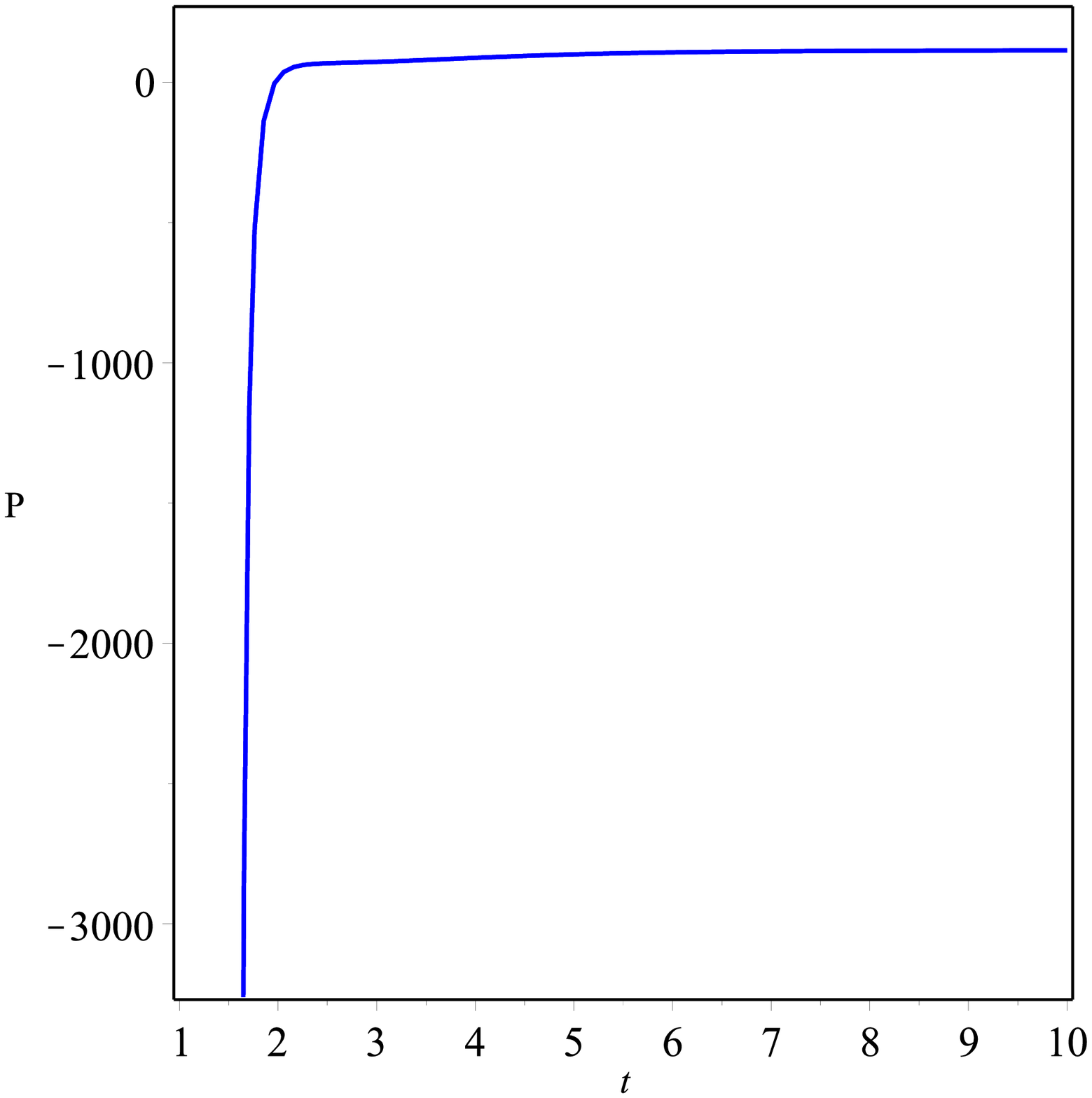} \vspace{12cm}\includegraphics{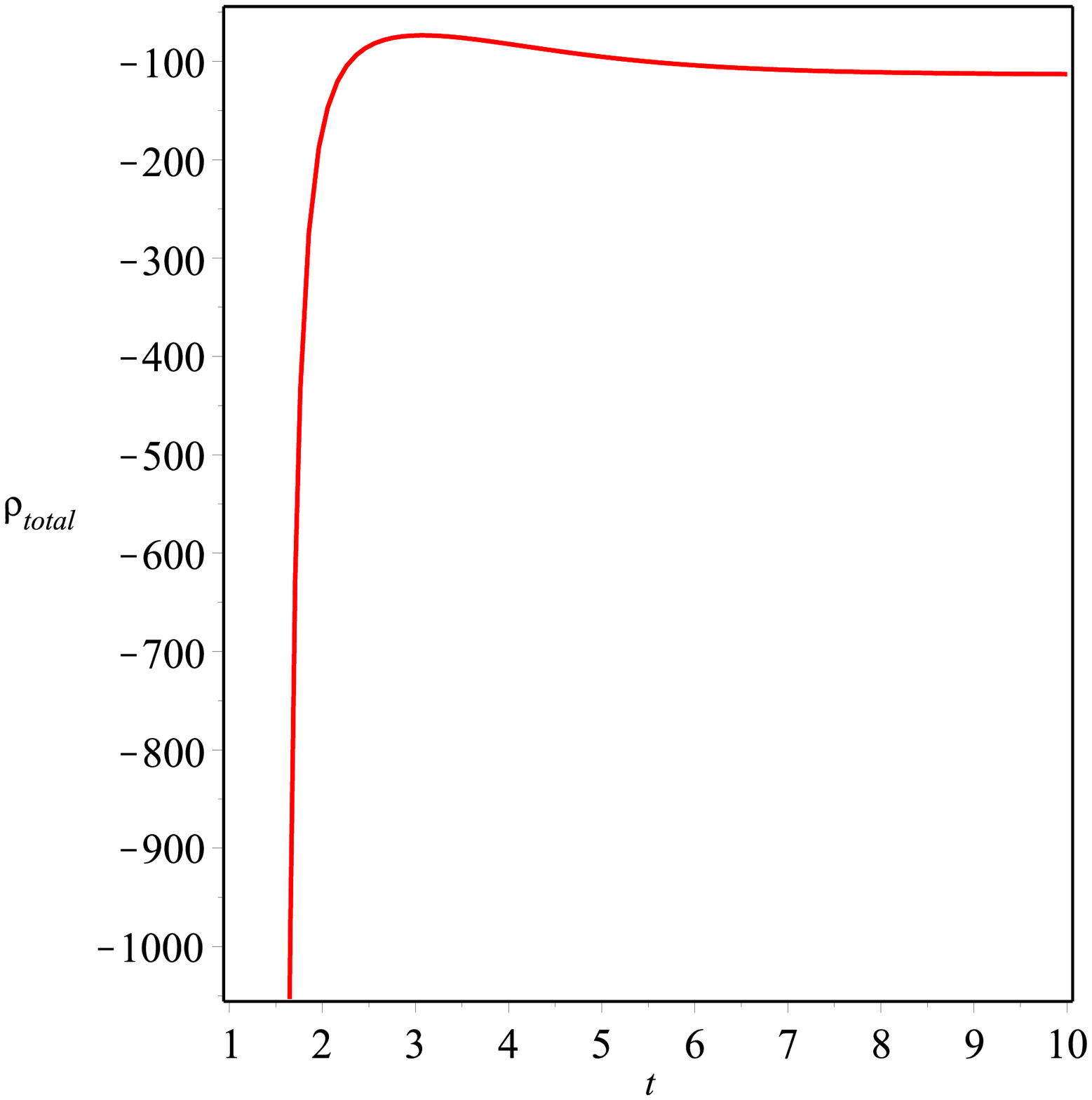}
\end{center}
\caption{\small {{\bf [a](Down).} Evolution of energy density
parameter,$\rho_{total}$, in terms of the cosmic time under the
logamediate scenario of the universe based on Nouicer GUP. {\bf
[b](UP).} Evolution of pressure parameter $P$ in terms of the cosmic
time under the logamediate scenario of the universe based on Nouicer
GUP . With values $n=1.0013$, $q=1$, $m_{p}=1$, $\alpha=2$,
$\mu=1.4$ (the values of the vertical axis are obtained according to
the values of the mentioned parameters.)}}
\end{figure}

In $Fig.11$, to investigate stability and instability, the behavior
of the square of the speed of sound,$\nu_{s}^2$, is plotted in terms
of cosmic time. According to the diagram $\nu_{s}^2$, in this part
the behavior of $\nu_{s}^2$ is divided into two step. In the first
step,$\nu_{s}^2$ has neither increasing nor decreasing
behavior(parabolic behavior) and is always at a positive level
($\nu_{s}^2>0$), which indicates the stability of the NADE model
under the logamediate scenario based on the Nouicer GUP, but in the
second step,$\nu_{s}^2$ has an increasing behavior and is always at
a negative level ($\nu_{s}^2<0$), which indicates the instability of
the NADE model under the logamediate scenario based on the Nouicer
GUP.
\begin{figure}[h]
\begin{center}\includegraphics{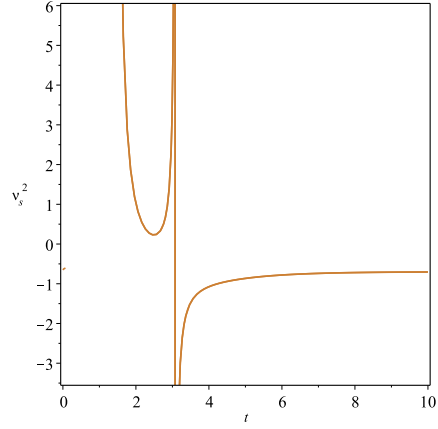} \vspace{6.5cm}
\end{center}
 \caption{\small {Evolution of the sound speed squares in terms of
the cosmic time to check the stability or instability of the dark
energy model under the logamediate scenario of the universe based on
Nouicer GUP. With values $n=1.0013$, $q=1$, $m_{p}=1$, $Q=0.05$,
$\alpha=2$, $\mu=1.4$ (the values of the vertical axis are obtained
according to the values of the mentioned parameters.)}}
\end{figure}

\subsection{Emergent scenario based on $GUP^*$}
According to the scale factor defined in the emergent scenario ($Eq.
56$), and its placement in the general expression of $Eq.30$
(equation of conformal time obtained in $Eq.57$) and also the
placement of $Eq.57$ in $Eq.52$(related to the energy density
$GUP^*$ in the previous section) energy density in this part is
obtained as follows:
\begin{equation}
\rho_{G4}=-\frac{9n^2m_{p}^2\hbar
Aa_{0}m(1+Be^{-At})^{-m}(B+e^{At})^m}{2n^2m_{p}^2\,_{2}F_{1}[m,m,1+m,-Be^{-At}]-q^2}
\end{equation}
Using $Eq.23$ and the Hubble parameter defined in the emergent
scenario ($Eq.60$),the new agegraphic dark energy pressure in this
part is obtained as follows:
\begin{equation}
p_{G4}=\frac{-Q-\dot{\rho_{G4}}}{3H}-\rho_{G4}
\end{equation}
According to $Eq.81$,$Eq.82$ and $Eq.59$,the state parameter
equation,$\omega_{total}$, and the square of the speed of
sound,$\nu_{s}^2$,are:
\begin{equation}
\omega_{total}=\frac{p_{G4}}{\rho_{G4}+\rho_{m1}}
\end{equation}
\begin{equation}
\nu_{s}^2=\frac{\dot{p_{G4}}}{\dot{\rho_{G4}}+\dot{\rho_{m1}}}
\end{equation}
In $Fig.12$, the equation of state parameter equation is plotted
under the emergent scenario based on $GUP^*$. Unlike the emergent
scenario in the Nouicer GUP,the emergent scenario based on $GUP^*$
describes the evolution of the universe in the far past, present,
and the far future times. In fact, the behavior of $\omega_{total}$
in this part is the same as the behavior of $\omega_{total}$ in the
emergent scenario based on Nouicer GUP, with the difference that the
evolution of the universe in this part, in addition to the far past,
is also generalized to the present and the far future. According to
$Fig.12$, in the mode of interaction with the evolution of the
universe, the value of $\omega_{total}$ decreases and in the present
and the far future time, the value of $\omega_{total}$ remain
constant, which means that the universe in the far past experienced
a period of accelerated expansion and with the evolution of the
universe in the present and the far future of the universe expand at
a constant rate, and in  non-interactive mode at all times, the
universe does not have accelerated expansion. According to
$\omega_{total}>-1$, so the universe has a quintessence-like
behavior in this part.
\begin{figure}[h]
\begin{center}\includegraphics{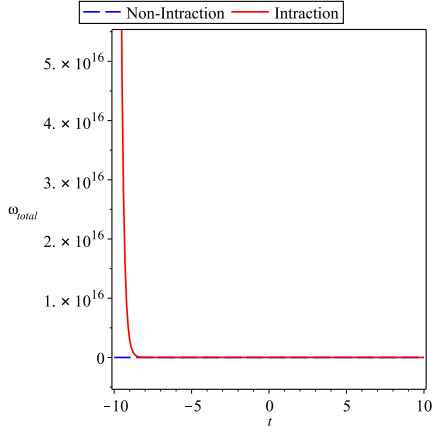} \vspace{6.5cm}
\end{center}
 \caption{\small {Evolution of state parameter equation in terms of
the cosmic time under the emergent scenario of the universe based on
$GUP^*$. With values $n=1.2$, $q=1$, $m_{p}=1$, $Q=0.05$,
$a_{0}=0.12$, $B=2.3$, $A=5.6$, $m=2$, $\hbar=2$ (the values of the
vertical axis are obtained according to the values of the mentioned
parameters.)}}
\end{figure}

In $Fig.13.a$ and $Fig.13.b$, the behavior of the $\rho_{total}$ and
$P$ parameters are plotted in terms of cosmic time. In these two
mode diagrams, interaction and non-interaction behave the same
exactly. With passing of time and evolution of the universe, the
value of $\rho_{total}$ is constant and at a certain time in the far
future,its value decreases rapidly, and the parameter $P$ is
constant with the evolution of the universe and passing of time, at
a certain time in the far future, its value increases rapidly.
\begin{figure}[h]
\begin{center}\includegraphics{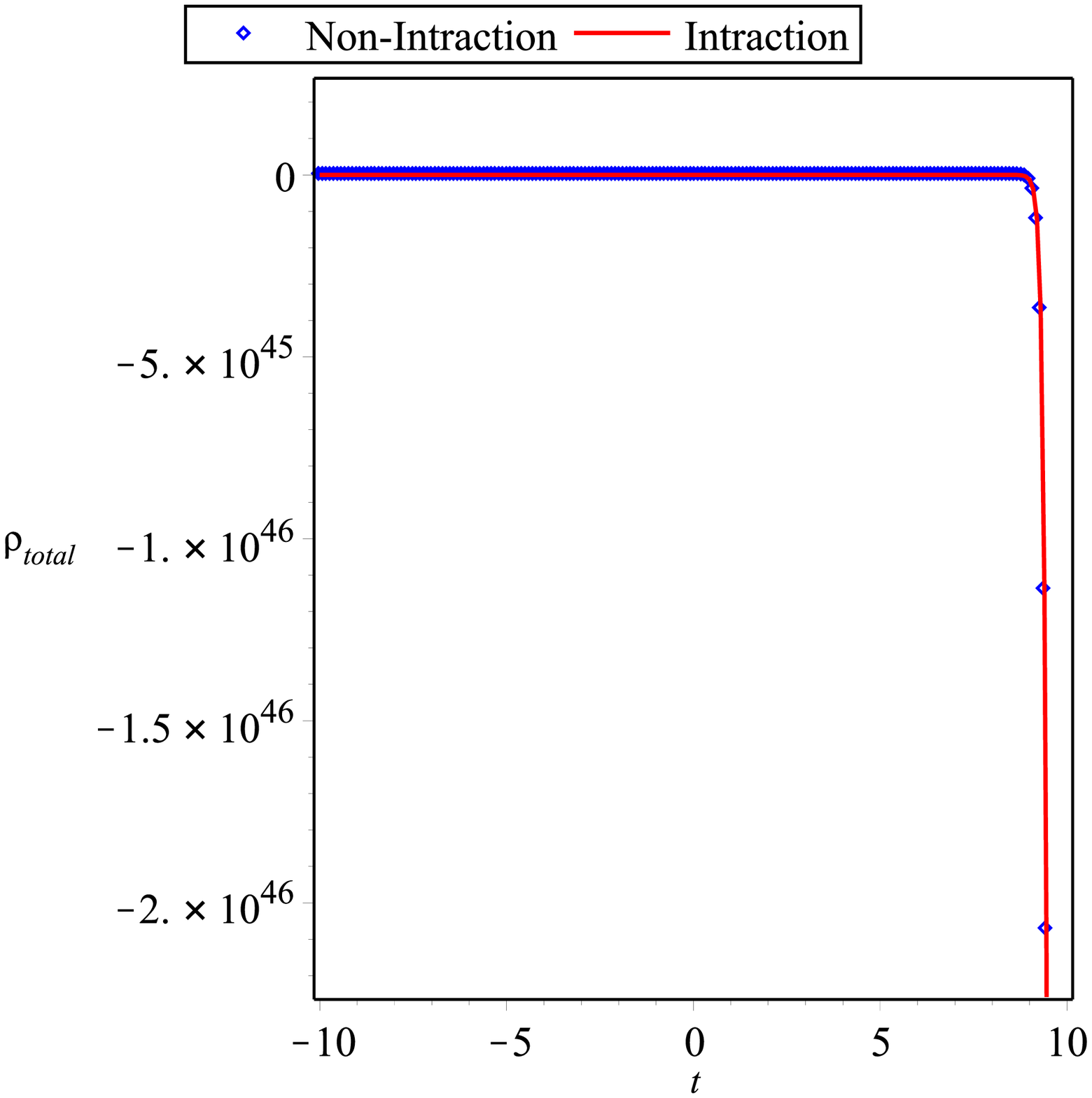} \vspace{12cm}\includegraphics{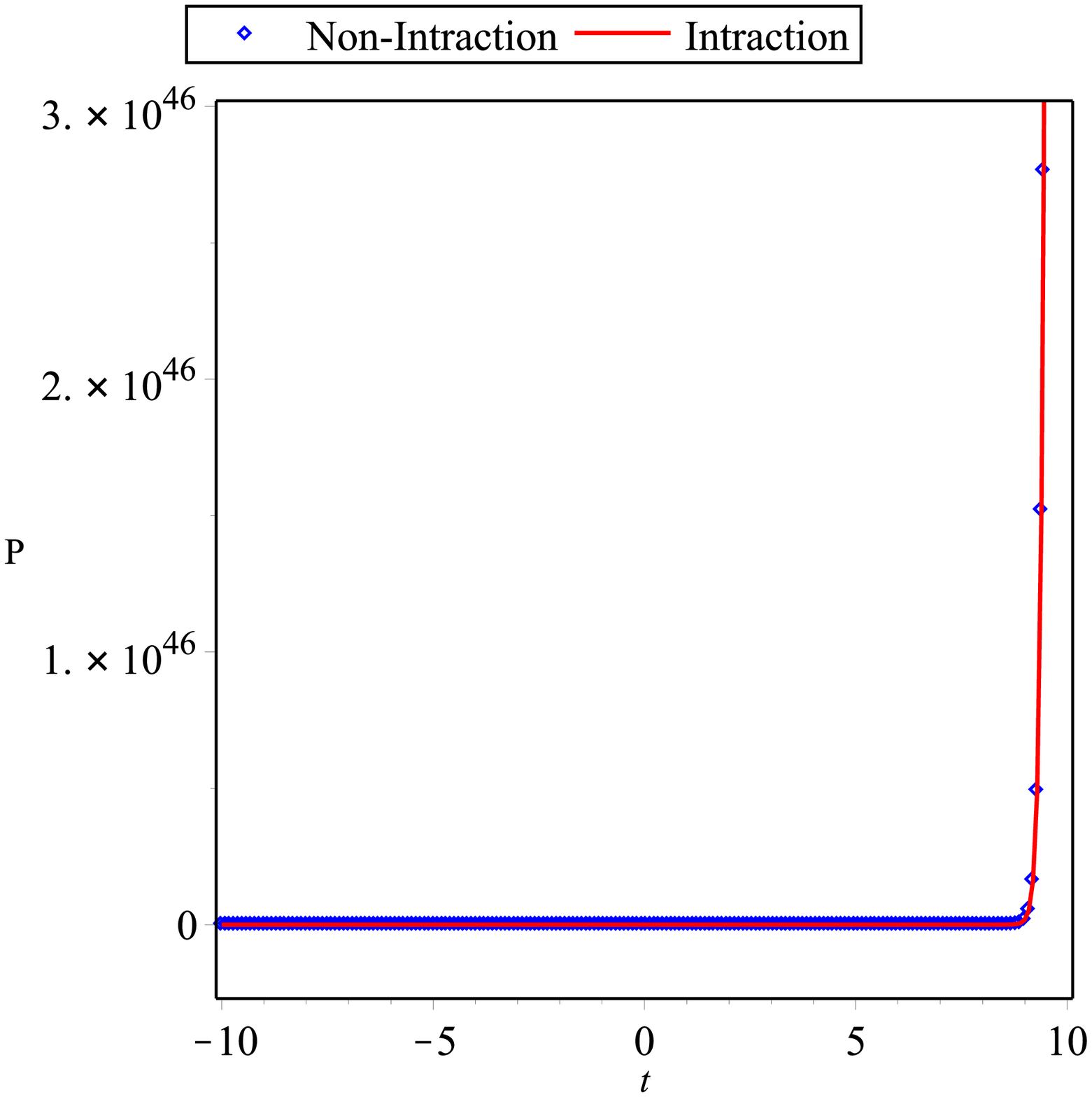}
\end{center}
\caption{\small {{\bf [a](Up).} Evolution of energy density
parameter,$\rho_{total}$, in terms of the cosmic time under the
emergent scenario of the universe based on $GUP^*$. {\bf [b](Down).}
Evolution of pressure parameter $P$ in terms of the cosmic time
under the emergent scenario of the universe based on $GUP^*$. With
values $n=1.2$, $q=1$ ,$m_{p}=1$, $Q=0.05$, $a_{0}=0.12$, $B=2.3$,
$A=5.6$, $m=2$, $\hbar=2$ (the values of the vertical axis are
obtained according to the values of the mentioned parameters.)}}
\end{figure}

In $Fig.14$,the sound speed squared behavior in both interaction and
non-interaction modes is the same as the behavior of
$\omega_{total}$, and $\nu_{s}^2$ has a decreasing behavior in this
part and always remains positive level($\nu_{s}^2>0$), which
indicates The stability of the NADE model under emergent scenario
based on $GUP^{*}$.
\begin{figure}[h]
\begin{center}\includegraphics{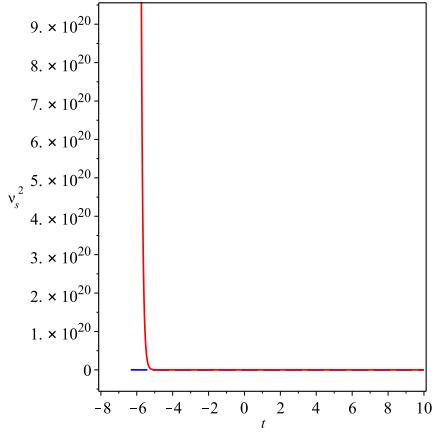} \vspace{6.5cm}
\end{center}
 \caption{\small {Evolution of the sound speed squares in terms of
the cosmic time to check the stability or instability of the dark
energy model under the emergent scenario of the universe based on
$GUP^*$. With values $n=1.2$, $q=1$, $m_{p}=1$, $Q=0.05$,
$a_{0}=0.12$, $B=2.3$, $A=5.6$, $m=2$, $\hbar=2$ (the values of the
vertical axis are obtained according to the values of the mentioned
parameters.)}}
\end{figure}

\subsection{Intermediate scenario based on $GUP^*$}
According to the scale factor defined in the intermediate scenario
($Eq.64$), and its placement in the general expression of $Eq.30$
(equation of conformal time obtained in $Eq.65$) and also the
placement of $Eq.65$ in $Eq.52$(related to the energy density
$GUP^*$ in the previous section) energy density in this part is
obtained as follows:
\begin{equation}
\rho_{G5}=\frac{-9n^6m_{p}^6\hbar\beta\lambda^{\frac{1}{\beta}}}{2(n^2m_{p}^2\Gamma[\frac{1}{\beta},\lambda
t^{\beta}]-q^2)}
\end{equation}
Using $Eq.23$ and the Hubble parameter defined in the intermediate
scenario($Eq.68$),the new agegraphic dark energy pressure in this
part is obtained as follows:
\begin{equation}
p_{G5}=\frac{Q-\dot{\rho_{G5}}}{3H}-\rho_{G5}
\end{equation}
According to $Eq.85$,$Eq.86$ and $Eq.67$, the parameter of state
equation,$\omega_{total}$,and the square of the speed of
sound,$v_{s}^2$,are:
\begin{equation}
\omega_{total}=\frac{p_{G5}}{\rho_{G5}+\rho_{m2}}
\end{equation}
\begin{equation}
\nu_{s}^2=\frac{\dot{p_{G5}}}{\dot{\rho_{G5}}+\dot{\rho_{m2}}}
\end{equation}
In $Fig.15$, the behavior of the state parameter equation in the
intermediate scenario based on $GUP^*$ is plotted. The intermediate
scenario based on $GUP^*$ is only able to describe the universe in
the present and the far future time and is not able to describe the
evolution of the universe in the far past time, and also in this
part, interaction and non-interaction modes behave exactly the same.
According to $Fig.15$,the evolution behavior of the universe is
divided into two steps; In the first step, with the evolution of the
universe and the passing of time, the value of $\omega_{total}$
decreases, which means that the universe has an accelerated
expansion, and in the second step, the value of $\omega_{total}$
decreases and then equals a constant value of $\omega_{total}=-1$,
which means the universe in The second step experiences an
accelerated expansion and then continues to expand at a constant
rate. In the first step, because $\omega_{total}<-1$, the universe
behaves the phantom-like, and in the second step, first because of
$\omega_{total}>-1$, the universe behaves the quintessence-like, and
then because of $\omega_{total}=-1$ behaves like the cosmological
constant.
\begin{figure}[h]
\begin{center}\includegraphics{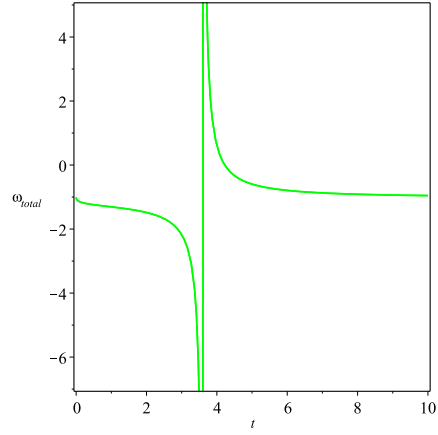} \vspace{6.5cm}
\end{center}
 \caption{\small {Evolution of state parameter equation in terms of
the cosmic time under the intermediate scenario of the universe
based on $GUP^*$. With values $n=2$ ,$q=1$, $m_{p}=1$, $Q=0.05$,
$\lambda=1.9$, $\beta=0.5$, $\hbar=2$ (the values of the vertical
axis are obtained according to the values of the mentioned
parameters.)}}
\end{figure}

In $Fig.16.a$ and $Fig.16.b$, the behavior of the parameters
$\rho_{total}$ and $P$ are plotted in terms of cosmic time, and both
interaction and non-interaction modes behave the same. With
evolution of the universe and passing of time, the value of
$\rho_{total}$ decreases in both steps. The value of the $P$
parameter increases rapidly in the first step and then decreases
rapidly in the second step.

\begin{figure}[h]
\begin{center}\includegraphics{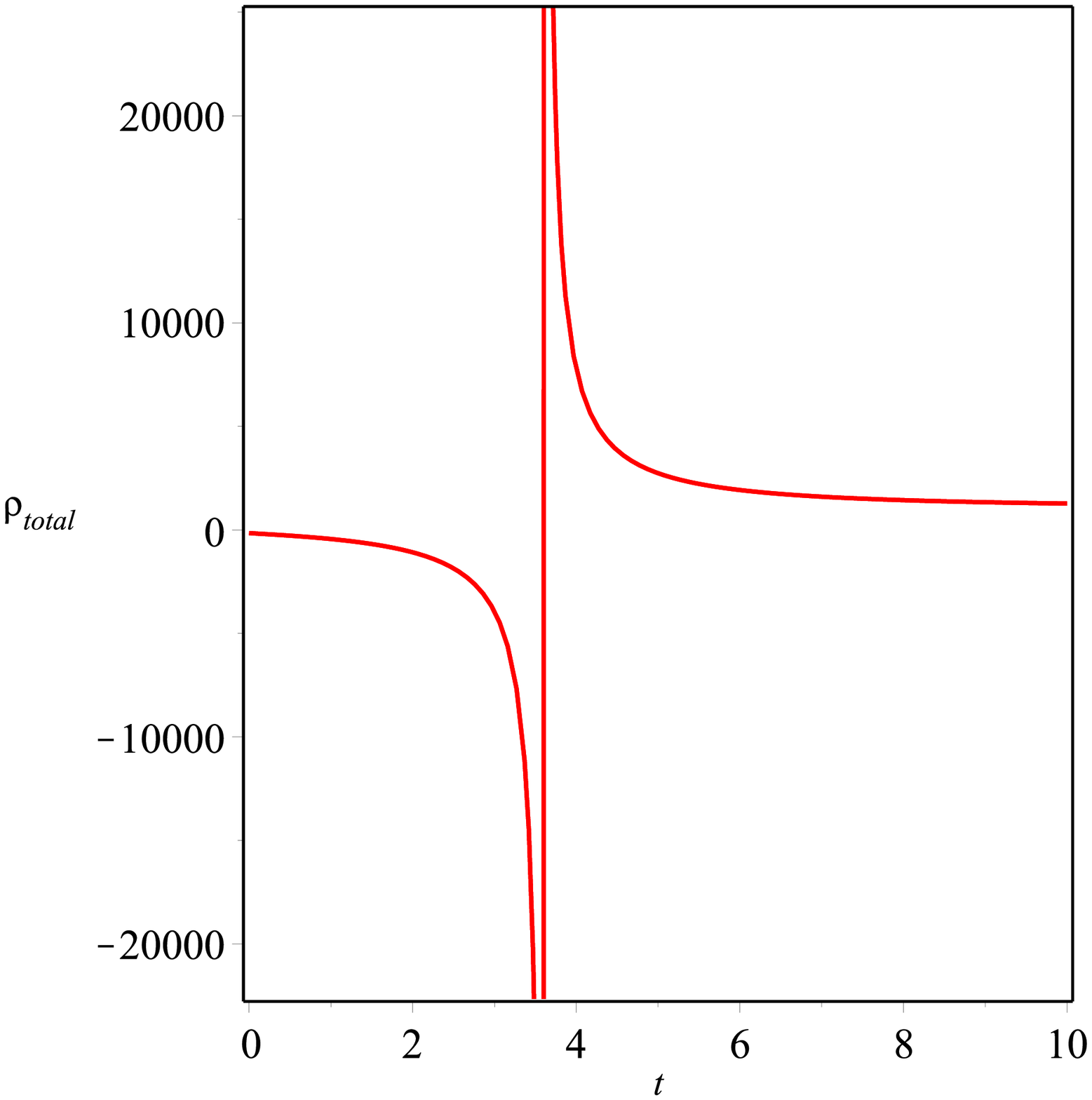} \vspace{12cm}\includegraphics{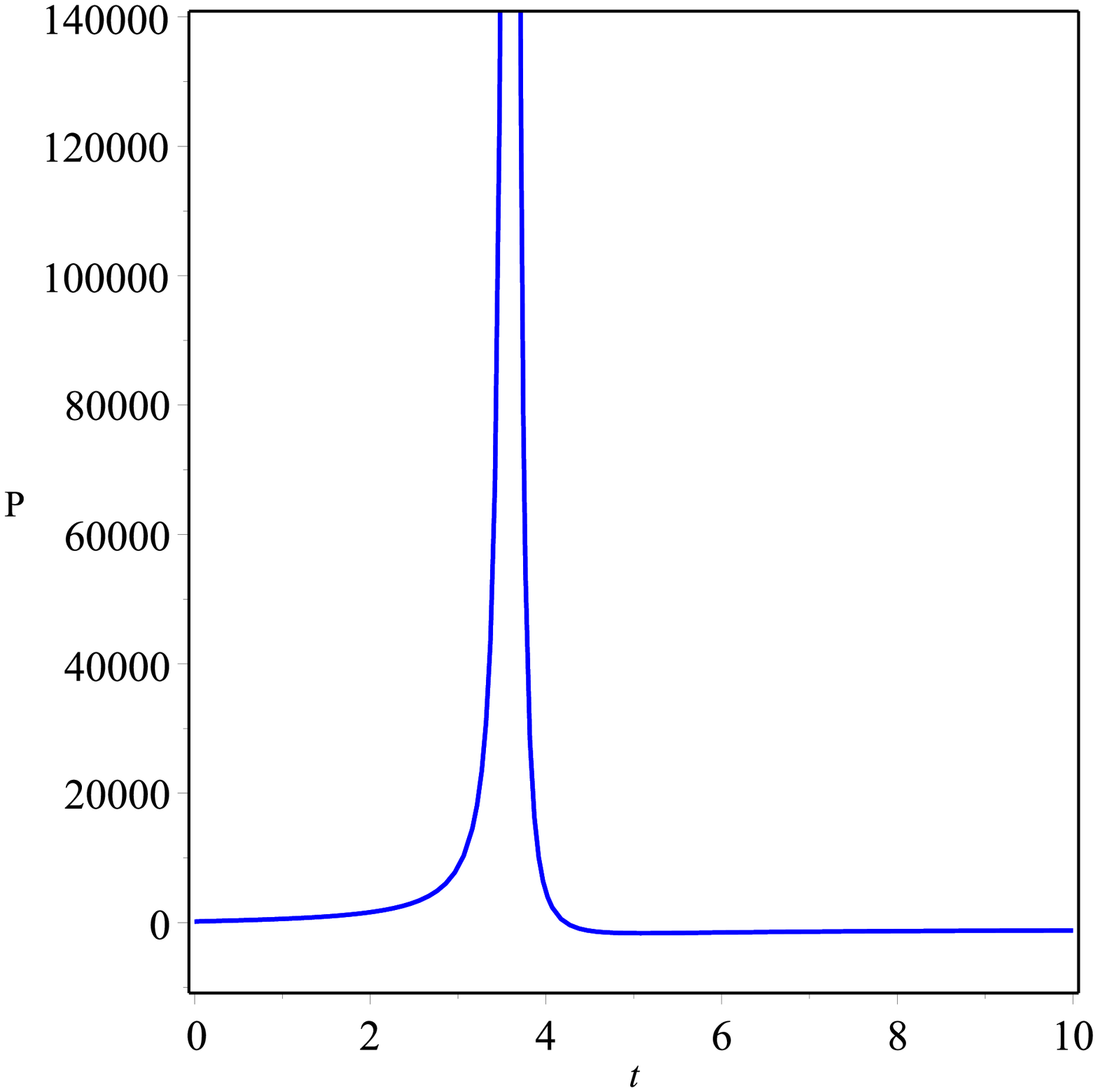}
\end{center}
\caption{\small {{\bf [a](Up).} Evolution of energy density
parameter,$\rho_{total}$, in terms of the cosmic time under the
intermediate scenario of the universe based on $GUP^*$. {\bf
[b](Down).} Evolution of pressure parameter $P$ in terms of the
cosmic time under the intermediate scenario of the universe based on
$GUP^*$. With values $n=2$, $q=1$, $m_{p}=1$, $Q=0.05$,
$\lambda=1.9$, $\beta=0.5$, $\hbar=2$ (the values of the vertical
axis are obtained according to the values of the mentioned
parameters.)}}
\end{figure}

$Fig.17$ shows the sound speed squared behavior in terms of cosmic
time. According to $Fig.17$, in the first step of the evolution of
the universe in this part,$\nu_{s}^2$ has a parabolic behavior and
is always at a negative level ($\nu_{s}^2<0$), which indicates the
instability of the NADE model, and in the second step of evolution,
$\nu_{s}^2$, has a decreasing behavior and always remains at a
positive level, indicating the stability of the NADE model under the
intermediate scenario based on $GUP^*$.
\begin{figure}[h]
\begin{center}\includegraphics{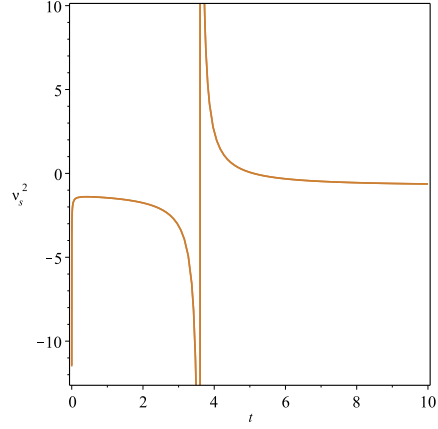} \vspace{6.5cm}
\end{center}
 \caption{\small {Evolution of the sound speed squares in terms of
the cosmic time to check the stability or instability of the dark
energy model under the intermediate scenario of the universe based
on $GUP^*$. With values $n=2$, $q=1$, $m_{p}=1$, $Q=0.05$,
$\lambda=1.9$, $\beta=0.5$, $\hbar=2$ (the values of the vertical
axis are obtained according to the values of the mentioned
parameters.)}}
\end{figure}

\subsection{Logamediate scenario based on $GUP^*$}
According to the scale factor defined in the intermediate scenario
($Eq.72$), and its placement in the general expression of $Eq.30$
(equation of conformal time obtained in $Eq.74$) and also the
placement of $Eq.74$ in $Eq.52$(related to the energy density
$GUP^*$ in the previous section) energy density in this part is
obtained as follows:
\begin{equation}
\rho_{G6}=\frac{9n^6m_{p}^6\hbar}{\frac{e^{\frac{1}{4\mu}}\sqrt{\pi}n^2m_{p}^2
erf(\frac{-1+2\mu lnt}{2\sqrt{\mu}})}{\sqrt{\mu}}-q^2}
\end{equation}
Using $Eq.23$ and the Hubble parameter defined in the logamediate
scenario ($Eq.77$),the new agegraphic dark energy pressure in this
part is obtained as follows:
\begin{equation}
p_{G6}=\frac{Q-\dot{\rho_{G6}}}{3H} - \rho_{G6}
\end{equation}
According to $Eq.89$,$Eq.90$ and $Eq.76$, the parameter of state
equation, $\omega_{total}$, and the square of the speed of sound,
$v_{s}^2$,are:
\begin{equation}
\omega_{G6}=\frac{p_{G6}}{\rho_{G6}+\rho_{m3}}
\end{equation}
\begin{equation}
\nu_{s}^2=\frac{\dot{p_{G6}}}{\dot{\rho_{G6}}+\dot{\rho_{m3}}}
\end{equation}
$Fig.18$ shows the behavior of the state parameter equation in terms
of cosmic time under the logamediate scenario based on $GUP^*$. The
logamediate scenario based on $GUP^*$ is only able to describe the
evolution of the universe in the present and the far future time,
and is not able to describe the evolution of the universe in the far
past time. Also,in this part,both interaction and non-interaction
modes behave exactly the same.According to $Fig.18$,the evolution of
the universe is divided into three steps; In the first step,the
value of $\omega_{total}$ increases rapidly, which means that the
universe has decelerated expansion. In the second
step,$\omega_{total}$ has a parabolic behavior,which means that the
universe will first have decelerated expansion and then accelerated
expansion. And in the third step, the value of $\omega_{total}$
decreases rapidly, which means that the universe has an accelerated
expansion. In the first step, first $\omega_{total}=-1$ and then its
value increases ($\omega_{total}>-1$) and this means that in the
first step, the universe first behaves like a cosmological constant
and then has a quintessence-like behavior. In the second step
$\omega_{total}<-1$ and this means that the universe will have
phantom-like behavior in this step. And in the third step,the
universe will have the same of the behavior of the first step.
Therefore, it will first have a quintessence-like behavior
($\omega_{total}>-1$) and then it will behave like a cosmological
constant ($\omega_{total}=-1$).
\begin{figure}[h]
\begin{center}\includegraphics{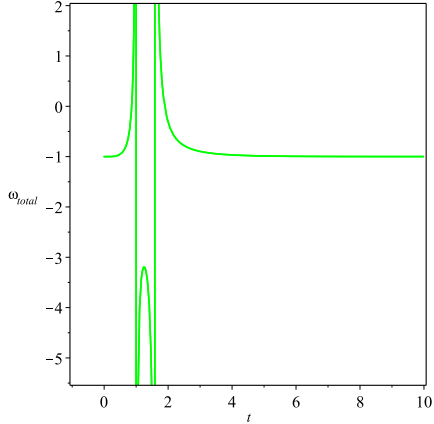} \vspace{6.5cm}
\end{center}
 \caption{\small {Evolution of state parameter equation in terms of
the cosmic time under the logamediate scenario of the universe based
on $GUP^*$. With values $n=1.0013$, $q=1$, $m_{p}=2$, $Q=0.05$,
$\alpha=2$, $\mu=1.4$, $\hbar=2$ (the values of the vertical axis
are obtained according to the values of the mentioned parameters.)}}
\end{figure}

In $Fig.19.a$ and $Fig.19.b$, the behavior of the parameters
$\rho_{total}$ and $P$ are plotted in terms of cosmic time,
respectively. With evolution of the universe and passing of time,
the value of $\rho_{total}$ decreases in all the steps, and the
parameter $P$ decreases in the first and third steps, and in the
second step it has a parabolic behavior. In these diagrams,the two
modes of interaction and non-interaction have the same behavior.
\begin{figure}[h]
\begin{center}\includegraphics{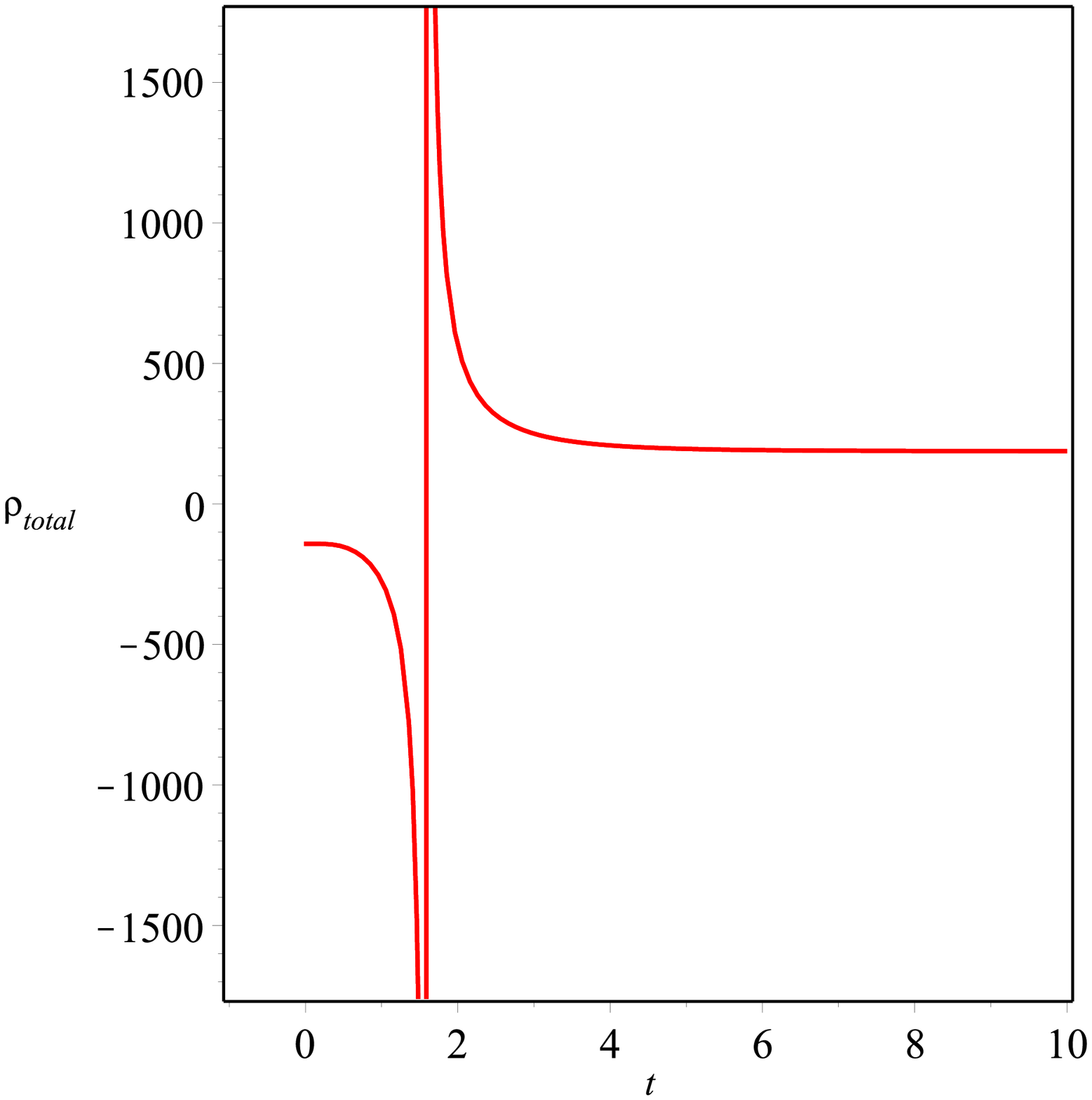} \vspace{12cm}\includegraphics{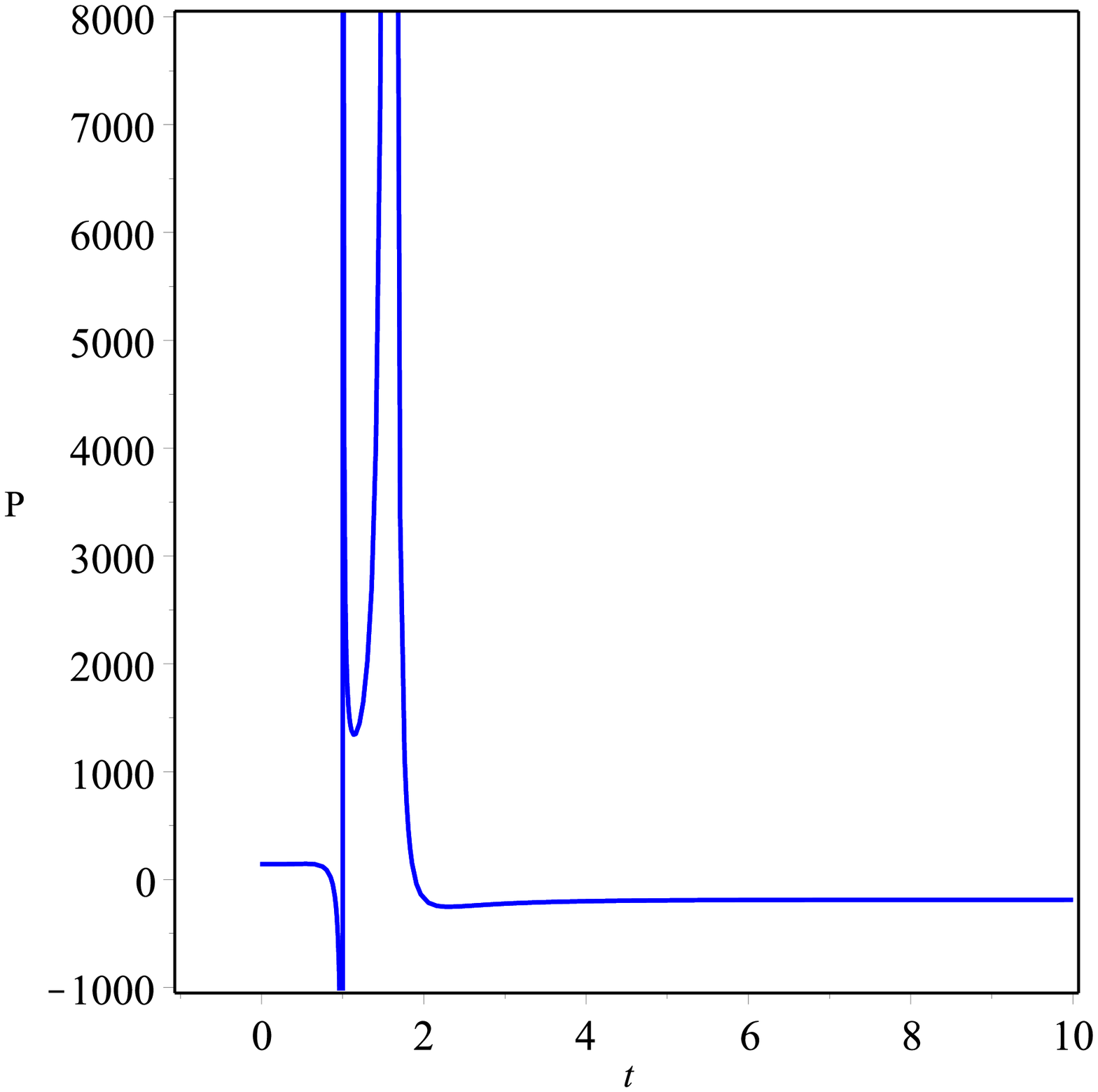}
\end{center}
\caption{\small {{\bf [a](Up).} Evolution of energy density
parameter,$\rho_{total}$, in terms of the cosmic time under the
logamediate scenario of the universe based on $GUP^*$. {\bf
[b](Down).} Evolution of pressure parameter $P$ in terms of the
cosmic time under the logamediate scenario of the universe based on
$GUP*$. With values $n=1.0013$, $q=1$, $m_{p}=2$, $\alpha=2$,
$\mu=1.4$, $\hbar=2$ (the values of the vertical axis are obtained
according to the values of the mentioned parameters.)}}
\end{figure}

In $Fig.20$, the  sound speed squared behavior is plotted in terms
of cosmic time. In the first step,$\nu_{s}^2$ has a parabolic
behavior and in the second step, because of $\nu_{s}^2$ decreases
with the evolution of the universe and tends towards
$\nu_{s}^2\rightarrow-\infty$, in the second step,the NADE model is
unstable. In the third step,$\nu_{s}^2$ has a decreasing behavior.
In the first and third steps of the evolution of the universe, in
the major part of the diagram,$\nu_{s}^2$ is positive and in a minor
part is negative, so by multiply from the positive and negative sign
and the result of the negative answer ($\nu_{s}^2<0$), we conclude
that $\nu_{s}^2$ is generally negative and this indicates the
instability of the NADE model under the logamediate scenario based
on $GUP^*$.
\begin{figure}[h]
\begin{center}\includegraphics{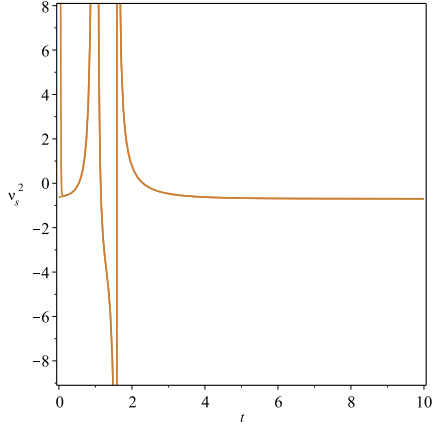} \vspace{6.5cm}
\end{center}
 \caption{\small {Evolution of the sound speed squares in terms of
the cosmic time to check the stability or instability of the dark
energy model under the logamediate scenario of the universe based on
$GUP^*$. With values $n=1.0013$, $q=1$, $m_{p}=2$, $Q=0.05$,
$\alpha=2$, $\mu=1.4$, $\hbar=2$ (the values of the vertical axis
are obtained according to the values of the mentioned parameters.)}}
\end{figure}
In Tables $2 , 3$ a summary of what is considered in Section 4 is
summarized.\\

\section{Conclusion}
The model(NADE) is investigated based on three types of GUP defined
in [56] (KMM, Nouicer and $GUP^*$) under three scenarios with
different scale factors(Emergent, Intermediate and Logamediate). Due
to the extensive work already done for the NADE model based on the
KMM GUP [see 20, 46, 49, 50, 57, 66], we decided to describe this
NADE model based on the Nouicer and $GUP^*$. In the second
section,the types of dark energy models (ADE) are reviewed and also
in the first subsection of the third section,the NADE model based on
KMM GUP is reviewed. In the second subsection of the third section
is defined the NADE model based on the Nouicer GUP. We concluded
that in this subsection, with the evolution of the universe and the
passing of time, the universe has an accelerated expansion, and the
reason is to neutralize the effects of the equations of evolution of
energy and dark matter($\Omega_{G}$ and $\Omega_{m}$) and reduce the
equation of state parameter ($\omega_{G}$). On the other hand, it
was concluded that in the far past,in the NADE model based on
Nouicer GUP, both matter and radiation are dominant, and with the
evolution of the universe, their effect on expansion decreases, and
instead the value of $\omega_{G}$ decreases and causes accelerated
expansion of the universe. In the third subsection of the third
section, the NADE model is defined based on $GUP^*$. In this
subsection,radiation ($\Omega_{G}$),matter ($\Omega_{m}$) and state
equation parameter ($\omega_{G}$) are all dominated in the far past
and far future times, which means that in these times the universe
has an acceleration expansion. In the present time, none of these
parameters are dominated,which all of these results are summarized
in $Table.1$. In Section $4$, the behavior and evolution of the
universe in the NADE model were investigated based on three
emergent, intermediate and logamediate scenarios based on KMM,
Nouicer and $GUP^*$. In each of the scenarios and GUPs, the behavior
of the  parameter  of state equation $\omega_{total}$, energy
density,$\rho_{total}$, pressure,$P$,and sound speed
squared,$\nu_{s}^2$ were investigated. According to [20, 46],the KMM
GUP is done and we just reviewed it in $Table.2$. But the Nouicer
GUP and $GUP^*$ show different behaviors than the KMM GUP of the
universe. In the emergent scenarios based on Nouicer GUP and $GUP^*$
the universe shows quintessence-like behavior. In the intermediate
scenario based on Nouicer GUP,the universe shows quintessence-like
behavior and based on $GUP^*$  the universe shows
quintessence-like,phantom-like and cosmological constant behavior.
In the logamediate scenario based on Nouicer GUP the universe has
the cosmological constant behavior in the present time and the
quintessence-like in the far future time. In the logamediate
scenario,based on $GUP^*$,the behavior and evolution of the universe
is limited to three steps, in the first and third steps of
evolution, the universe has a constant cosmological and
quintessence-like behavior, but in the second step of evolution,the
universe shows phantom-like behavior. The NADE model based on KMM
GUP is unstable under all scenarios but the NADE model based on
Nouicer GUP is stable under emergent and intermediate scenarios, and
under the logamediate scenario is stable in the first step of
evolution and unstable in the second step of evolution. In
$GUP^*$,under the emergent scenario,the NADE model is stable, and
under the intermediate scenario,the model is unstable in the first
step of evolution, also the model is stable in the second step of
evolution, and the NADE model based on $GUP^*$ is unstable under the
logamediate scenario. In general,the behavior and evolution of the
universe in the types of GUPs and under all three mentioned
scenarios in Tables $2,3$ are summarized.\\
\onecolumn
\begin{table}[p]
\begin{center}
\caption{summary table of results.} \vspace{0.1 cm}
\begin{tabular}{|c|c|p{3cm}|p{2cm}|p{2cm}|p{2cm}|p{2.1cm}|}
  \hline
  {\bf GUP} &{\bf Scenario} & {\bf $\omega_{total}$} & {\bf $\rho_{total}$} & {\bf P} & {\bf $\upsilon_s^2$} & {\bf Description} \\
  \hline
  KMM & Emergent & {\tiny Increasing behavior $\omega_{total}<-1$ Phantom-like
  behavior}
 & Upward $\uparrow$
 & Downward $\downarrow$ & $\upsilon_s^2< 0$ & {\tiny The model is instable.} \\
  \hline
  KMM & Intermediate & {\tiny Increasing behavior $\omega_{total}<-1$ Phantom-like
  behavior}
 & Upward $\uparrow$ & Downward $\downarrow$ & $\upsilon_s^2< 0$  & {\tiny The model is instable.} \\
  \hline
  KMM & Logamediate & {\tiny Increasing and decreasing behavior $\omega_{total}>-1$ quintessence-like
  behavior}
 & Downward $\downarrow$ & Upward $\uparrow$ & $\upsilon_s^2< 0$  & {\tiny The model is instable.} \\
  \hline
  Nouice & Emergent & {\tiny The Description of the universe evolution in the far past. Decreasing behavior in the interaction mode and be constant in
  the non-interaction mode. $\omega_{total}>-1$ Quintessence like
  behavior}
 & Downward $\downarrow$ & Downward $\downarrow$ & $\upsilon_s^2> 0$  & {\tiny The model is stable.} \\
  \hline
  Nouice & Intermediate & {\tiny The Description of the universe evolution in the far past. Increasing behavior. $\omega_{total}>-1$ Quintessence like
  behavior}
 & Upward $\uparrow$ & Downward $\downarrow$ & $\upsilon_s^2> 0$  & {\tiny The model is stable.} \\
  \hline
  Nouice & Logamediate & {\tiny The Description of the universe evolution in the present and the far future times. Decreasing behavior.$\omega_{total}>-1$ For far future time and $\omega_{total}=-1$ For present time.
Quintessence-like behavior for the far future time and cosmological
constant behavior for the present time}
 & Upward $\uparrow$ & Upward $\uparrow$ & {\tiny First step $\upsilon_s^2> 0$ and second step $\upsilon_s^2< 0$ }  & {\tiny In the first step, the model is
stable and  in the second step, the model is instable.}
 \\
  \hline
\end{tabular}
\end{center}
\end{table}

\newpage

\begin{table}[p]
\begin{center}
\caption{summary table of results.} \vspace{0.1 cm}
\begin{tabular}{|c|c|p{3cm}|p{2cm}|p{2cm}|p{2cm}|p{2.1cm}|}
  \hline
  {\bf GUP} &{\bf Scenario} & {\bf $\omega_{total}$} & {\bf $\rho_{total}$} & {\bf P} & {\bf $\upsilon_s^2$} & {\bf Description} \\
  \hline
  $GUP^*$ & Emergent & {\tiny The Description of the universe evolution in all of the times. Decreasing behavior for the interaction mode and be constant behavior for non-interaction
  mode.$\omega_{total}>-1$ Quintessence-like behavior}
 & Downward $\downarrow$ & Upward $\uparrow$ & $\upsilon_s^2> 0$ &  {\tiny the model is stable.} \\
  \hline
  $GUP^*$ & Intermediate & {\tiny The Description of the universe evolution in the present and the far future times. Decreasing behavior. $\omega_{total}\leq-1$ in the first step And $\omega_{total}\geq-1$ in the second step, The phantom-like and cosmological constant behavior in the first step And The quintessence-like and cosmological constant behavior in the second
  step}
 & {\tiny Downward $\downarrow$ for both steps }& {\tiny Upward$\uparrow$ For the first step And Downward $\downarrow$ For the second
 step}
 & {\tiny First step $\upsilon_s^2< 0$ and second step $\upsilon_s^2> 0$ } & {\tiny In the first step the model is instable and in the second step the model is stable.} \\
  \hline
  $GUP^*$ & Logamediate & {\tiny The Description of the universe evolution in the present and the far future times. Increasing behavior in the first step , parabolic behavior in the second step and decreasing behavior in the third step. $\omega_{total}\geq-1$ in the first and third steps and $\omega_{total}<-1$ In the second
step. Quintessence-like and cosmological constant behavior in the
first and third steps and Phantom-like behavior in the second step}
 & Downward $\downarrow$ & {\tiny Downward $\downarrow$ For the first and third steps of
 evolution. Parabolic $\updownarrow$ For the second step of
 evolution}
  & {\tiny In generally and multiplying the sign of $\upsilon_s^2$ in all of step: $\upsilon_s^2 <0$}
& {\tiny The model is instable.} \\
  \hline
\end{tabular}
\end{center}
\end{table}

\twocolumn

\end{document}